# High-Efficiency Three-Wave and Four-Wave Phonon Mixing Via Electron-Mediated Nonlinearity in Semiconductor-Piezoelectric Heterostructures

Lisa Hackett,[1] Matthew Koppa,[1] Brandon Smith,[1] Michael Miller,[1] Steven Santillan,[1] Scott Weatherred,[1] Shawn Arterburn,[1] Thomas A. Friedmann,[1] Nils Otterstrom,[1] and Matt Eichenfield[1,2]

1. Microsystems Engineering, Science, and Applications, Sandia National Laboratories, Albuquerque, NM 87123, USA
2. College of Optical Sciences, University of Arizona, Tucson, AZ, 85719, USA

Correspondence: Matt Eichenfield (eichenfield@arizona.edu)

**Abstract**
We show that phononic frequency conversion can be enhanced by orders of magnitude in piezoelectric systems by heterogeneous integration of high-mobility semiconductor films. A lithium niobate and indium gallium arsenide heterostructure is utilized to demonstrate efficient three-wave mixing processes at microwave frequencies, including $(16\pm6)$% phononic power conversion efficiency for sum-frequency generation and $(1.0\pm0.1)$% phononic power conversion efficiency for difference-frequency generation, as well as the most efficient degenerate four-wave phononic mixing to date. We present a theoretical model that accurately predicts the sum-frequency and difference-frequency generation processes and we show that the conversion efficiency can be further enhanced by the application of semiconductor bias fields. Laser Doppler vibrometry is then applied to examine many three-wave and four-wave mixing processes simultaneously in the same device. Through the use of our developed model, we show that these nonlinearities can be enhanced far beyond what is demonstrated here by confining phonons to smaller dimensions in waveguides and optimizing semiconductor material properties or using 2D semiconductors.

# Introduction

Nonlinear acoustic interactions have the potential to completely transform the way we think of and use phonons in the solid state, just as nonlinear optical interactions have done for photons.[1-3] In the classical domain, nonlinear acoustic frequency conversion can be used to create frequency conversion devices, such as mixers and correlators,[4-7] that can greatly enhance radio frequency signal processing or even form the basis for all-acoustic radio frequency signal processing on a chip.[8] In the quantum domain, strong nonlinear multi-phonon interactions could allow for quantum phononic processes such as squeezing, parametric amplification, state-preserving frequency conversion, and phononic quantum logic to be harnessed in ways analogous to their use in quantum photonics.[9-11] Nonlinear phononic processes are also known to limit the coherence of phonons in materials in important regimes.[12,13] Thus the ability to understand, modify, and control those nonlinearities can offer deep insight into or enable the control of heat transport,[14,15] the kinetics and dynamics of charge carriers,[16-18] and other important solid state physical processes. However, much like optical nonlinearities, the strength and utility of acoustic nonlinearities in solids are entirely material dependent. While materials like lithium niobate do show some promise for nonlinear acoustics, with useful demonstrations of electro-acoustic effects,[19] nonlinear piezoelectric effects,[20] and three-wave[21,22] and four-wave mixing processes,[23-27] thus far strong acoustic nonlinearities that enable high-efficiency frequency conversion in solid state materials have not been identified.

In this work, we provide an alternative approach to achieving strong phononic nonlinearities by demonstrating efficient phononic mixing processes in piezoelectric acoustic systems by heterogeneous integration of semiconductor films. This enhancement to phononic frequency conversion efficiency occurs due to hybridization of the phonons with electronic waves in the semiconductor, with the degree of hybridization determined by the strength of the piezoelectric phonons' longitudinal electric field created in the semiconductor. Strong *electronic* nonlinearities then can mediate strong phononic nonlinearities. We use this acoustoelectrically enhanced frequency conversion efficiency to demonstrate highly efficient three-wave mixing processes at microwave frequencies, including a (16±6)% phononic power conversion efficiency for sum-frequency generation and (1.0±0.1)% phononic power conversion efficiency for difference-frequency generation, as well as the most efficient degenerate four-wave phononic mixing to date. We show the frequency conversion can be further enhanced by application of bias fields in the semiconductor. In the case of four-wave phononic mixing, the natural phase matching of the process enables us to directly experimentally show that the improvement to the phononic frequency conversion efficiency is from the enhanced nonlinear coefficient due to the presence of the semiconductor film. We present a theoretical model that accurately predicts the sum-frequency and difference-frequency generation processes. We then use laser Doppler vibrometry to examine many three-wave and four-wave mixing processes simultaneously in the same device, including the ones described above and others such as second and third harmonic generation. Finally, we use the theoretical model to show that these nonlinearities, and the resulting power conversion efficiency, can be greatly enhanced far beyond even what is demonstrated here by confining phonons to smaller dimensions in waveguide circuits and optimizing semiconductor material properties or using 2D semiconductor materials. This work paves the way for deterministic integration of semiconductor materials into piezoelectric acoustic materials and circuits for the purpose of producing, studying, controlling, and using strong phononic nonlinearities, which can be combined with other acoustoelectric effects to produce novel classes of phononic devices and materials systems.

**The acoustoelectric nonlinearity**
Here we study frequency conversion of collinear and co-propagating phonons mediated by an acoustoelectric nonlinearity in a semiconductor piezoelectric heterostructure, as illustrated in Fig. 1(a). As an illustrative example, a parametric three-wave mixing process is shown where phonons at frequencies $\omega_1$ and $\omega_2$ interact nonlinearly to generate a phonon at the sum frequency $\omega_3 = \omega_1 + \omega_2$. The phonons in this work are produced in modes of a planar piezoelectric waveguide, which endows the phonons with electric fields that extend evanescently above the surface of the piezoelectric material. A nonlinear interaction region is created by selectively patterning a semiconductor layer to be directly above the waveguide, where the phonons' electric fields couple to and become hybridized with acoustoelectrically generated propagating charge distributions in the semiconductor. The strong coupling and hybridization between the semiconductor charge and piezoelectric phonons allows the electronic nonlinearities to mediate nonlinear interactions between the phonons.

The system we utilize is a semiconductor piezoelectric heterostructure on a silicon substrate that allows independent control over the acoustic, electromechanical, and semiconducting material parameters that determine the linear and nonlinear acoustic response. The specific heterostructure utilized in this work consists of an indium gallium arsenide ($In_{0.53}Ga_{0.47}As$) semiconducting film with a thickness of approximately 50 nm on a 5 μm thick lithium niobate ($LiNbO_3$) piezoelectric layer. The $In_{0.53}Ga_{0.47}As$ is grown by metal-organic chemical vapor deposition within a $In_{0.53}Ga_{0.47}As$/indium phosphide (InP) epitaxial ternary heterostructure on an InP wafer. The entire epitaxial stack is lattice-matched to InP, enabling a low semiconductor defectivity and therefore high mobility, $\mu$. As discussed in detail in the Methods, the complete heterostructure is formed by first bonding the $In_{0.53}Ga_{0.47}As$/InP stack to a $LiNbO_3$-silicon substrate followed by removal of the InP substrate. The remaining epitaxial layers are patterned and etched to create the semiconductor nonlinear interaction region ($In_{0.53}Ga_{0.47}As$) and semiconductor mesa contacts to allow drift fields and currents to be applied to that layer (see Methods). $LiNbO_3$ is a highly anisotropic piezoelectric material where the electromechanical coupling coefficient, $k^2$, can be exceptionally large for specially chosen acoustic modes and directions. Here we target a mode with quasi-shear horizontal (quasi-$SH_0$) polarization that propagates along the $x$-direction in $y$-cut $LiNbO_3$ with a $k^2 \sim 18\%$. The modeled longitudinal electric field and displacement field for the quasi-$SH_0$ mode in our acoustoelectric heterostructure is shown in Fig. 1(b). As can be seen, the electric field value of the phonon mode extends above the surface of the piezoelectric material into the semiconductor and peaks roughly at the interface between the two, which enables the strong coupling between the phonon and semiconductor charge carriers.

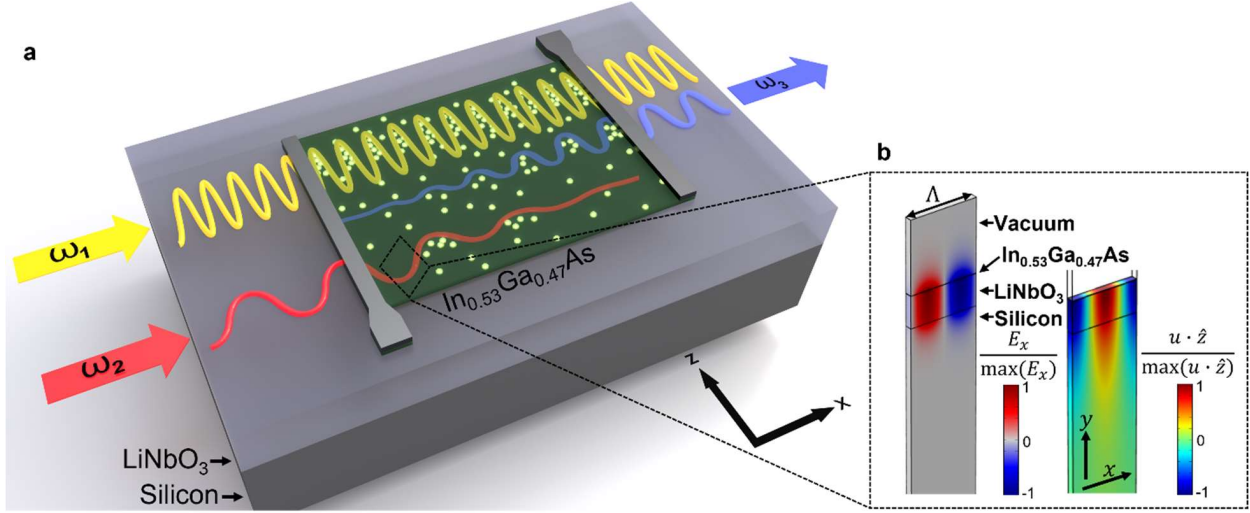

**Fig. 1: Phononic frequency mixing mediated by an acoustoelectric nonlinearity.** (a) Illustration of sum frequency generation with collinear and co-propagating piezoelectric phonons coupled to a semiconductor charge carrier system. The frequency conversion is mediated by an acoustoelectric nonlinearity in the heterostructure system. In the actual devices, the acoustic and electronic waves physically overlap. (b) Modeled longitudinal electric field ($E_x$) and z-component of the displacement field ($u \cdot \hat{z}$) for a guided piezoelectric acoustic wave with quasi-shear horizontal ($SH_0$) polarization in a In$_{0.53}$Ga$_{0.47}$As-LiNbO$_3$ film stack on a silicon substrate. The In$_{0.53}$Ga$_{0.47}$As thickness is 50 nm, the LiNbO$_3$ thickness is 5 μm, and the modeled acoustic wavelength, $\Lambda$, is $\Lambda = 16$ μm.

In the remainder of this Article, we first assess two parametric three-wave mixing processes: sum and difference frequency generation, which rely on a second-order acoustoelectric susceptibility, $X_{AE}^2$. We examine the power-dependent conversion efficiency in the absence of biasing electric fields in the semiconductor. We then show how the sum frequency conversion efficiency can be modified through an applied static electric field in the semiconductor. The study of frequency conversion is then extended to parametric four-wave mixing, utilizing the third-order acoustoelectric susceptibility, $X_{AE}^3$. We then demonstrate that laser Doppler vibrometry can be used to capture a broad range of nonlinear mixing processes in a single device and show that myriad nonlinear processes occur in these heterostructures that can be studied in future work. A model is developed to assess how to maximize the nonlinearities while simultaneously minimizing loss in these systems by proper choice of materials and geometry, and we show a path to achieve power conversion efficiencies significantly beyond what is demonstrated here for sum frequency generation.

**Sum and difference frequency generation**
Figure 2(a) shows the energy level diagram for sum frequency generation. Energy conservation requires that two phonons, one at $\omega_1$ and one at $\omega_2$, are annihilated to create a phonon at $\omega_3 = \omega_1 + \omega_2$. Figure 2(b) shows a microscope image of a device implemented in the semiconductor piezoelectric heterostructure to study this frequency conversion process. To generate and detect phonons, we utilize interdigital transducers that consist of interlaced metal electrodes on the piezoelectric surface. Two of these transducers are designed and fabricated to generate collinear and copropagating pump and signal phonons at $f_1 = \omega_1/2\pi$ and $f_2 = \omega_2/2\pi$, respectively.

These phonons propagate through a 250 μm long region of patterned In$_{0.53}$Ga$_{0.47}$As semiconductor film, which supports the acoustoelectric interaction. The output phonons then pass under an interdigital transducer designed to be resonant at $f_3 = \omega_3/2\pi = \omega_1/2\pi + \omega_2/2\pi$, which generates an electrical signal for detection. While several nonlinear processes may occur simultaneously in this single device, the output transducer is an efficient filter of $f_1$, $f_2$, and the converted frequencies outside of the passband centered at $f_3$. Electrodes on the In$_{0.53}$Ga$_{0.47}$As enable modification of the electrical boundary condition to the patterned semiconductor, which we show can modify the conversion efficiency. A plot of the measured electrical reflection, $S_{11}$, as a function of frequency for the three interdigital transducers used for phonon generation and detection in the system is shown in Fig. 2(c) (reflection measurement details are given in the Methods). When the input electrical signal frequency matches the designed resonance frequency of the transducer, a significant amount of power is radiated as piezoelectric phonons, as evidenced by a characteristic dip in the $S_{11}$ spectrum. As shown in Fig. 2(c), the measured resonance frequencies are $f_1 = 370$ MHz, $f_2 = 256.6$ MHz, and $f_3 = 626.6$ MHz.

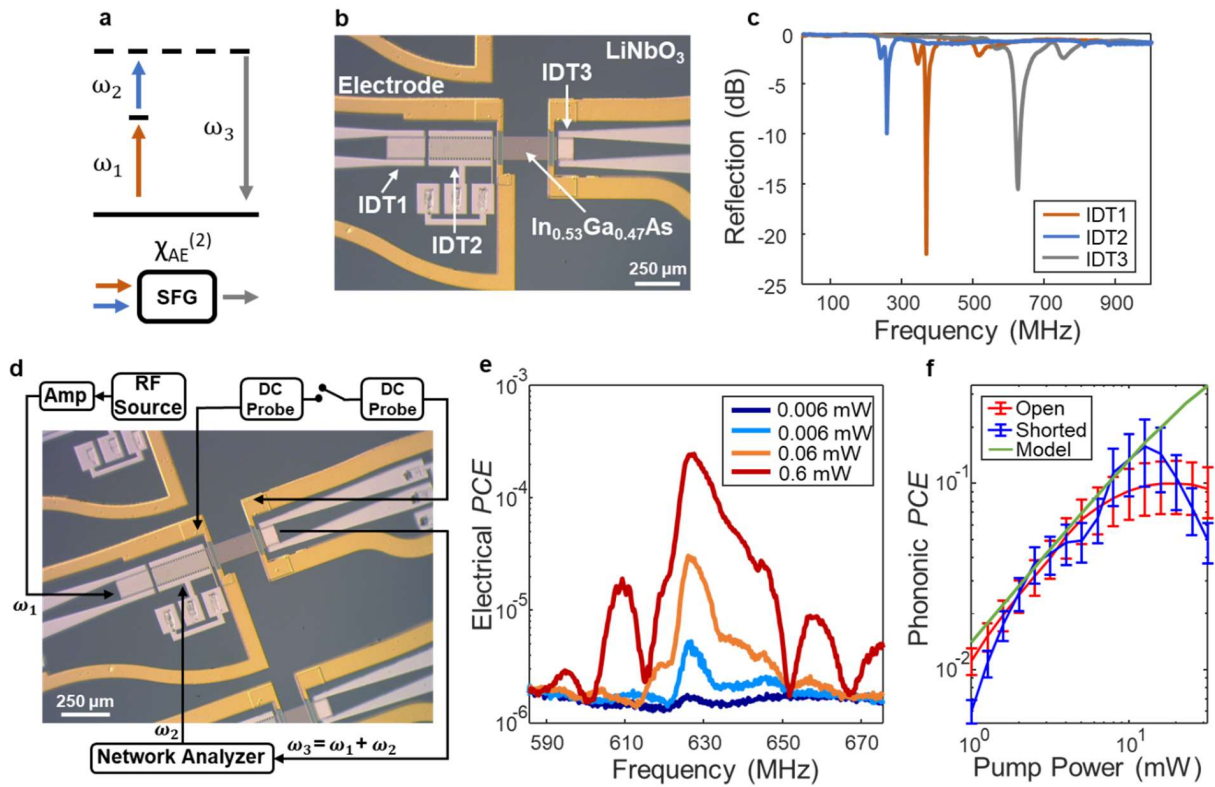

**Fig. 2: Sum frequency generation.** (a) Phonon energy level diagram for sum frequency generation. (b) Microscope image of physical system implementation in the In$_{0.53}$Ga$_{0.47}$As - LiNbO$_3$ semiconductor-piezoelectric heterostructure on a silicon substrate. Two input interdigital transducers (IDTs) launch the collinear and co-propagating pump and signal phonons, which propagate through a 250 μm long patterned In$_{0.53}$Ga$_{0.47}$As semiconductor on the LiNbO$_3$ surface that defines the acoustoelectric interaction region. An output IDT on the LiNbO$_3$ is designed to have a resonant frequency equal to the sum frequency. (c) Measured reflection as a function of frequency for the pump, signal, and sum frequency IDTs. (d) Schematic of the experimental setup to study sum frequency generation. The IDTs convert radio frequency electrical signals to radio frequency piezoelectric acoustic waves. The acoustoelectric interaction region, defined by the patterned semiconductor layer on the piezoelectric film, provides the nonlinearity. (e) Measured terminal power conversion efficiency (*PCE*) as a function of the detection frequency for different phononic pump powers. (f) Measured phononic power conversion efficiency (*PCE*) as a function of phononic pump power at a single detection frequency for open and shorted electrical contact to the $In_{0.53}Ga_{0.47}As$ semiconductor. The modeled values according to numerical integration of the nonlinear coupled-mode equations (Supplementary Note 3) are also shown.

Figure 2(d) shows an experimental schematic for the sum frequency generation power conversion efficiency measurements and additional details are given in the Methods section. A network analyzer is used as the electrical source for the transducer resonant at the signal frequency $f_2$ and as the electrical detector for the transducer resonant at the sum frequency $f_3$. An amplified radio frequency source is used as the electrical input to the interdigital transducer to generate the pump phonon at $f_1$. The network analyzer gives a direct measurement of the electrical power conversion efficiency from $f_2$ to $f_3$, which is the end-to-end total power conversion efficiency that includes the losses incurred during the transduction processes utilized for generation and detection of phonons. We separately fabricated acoustic delay line test structures that allowed us to measure the transducer conversion losses (see Supplementary Note 1). These measured conversion losses and the electrical power conversion efficiency then allow us to quantify the phononic power conversion efficiency.

Figure 2(e) shows a plot of the electrical power conversion efficiency as a function of the detection frequency for several different phononic pump powers. In this experiment, the pump frequency was fixed at $f_1 = 370$ MHz, the signal frequency was varied from $f_2 = 206.6$ MHz to $f_1 = 306.6$ MHz, and the detection frequency was modified to satisfy $f_3 = f_1 + f_2$ The phononic pump power is the pump power at the beginning of the patterned $In_{0.53}Ga_{0.47}As$ semiconductor that defines the acoustoelectric interaction region. As can be seen in Fig. 2(e), the electrical power conversion efficiency achieves a maximum value at the resonance frequency of the output transducer, increases with increasing pump power, and has a 3 dB fractional bandwidth of 1.2%, which is limited by the transducer bandwidth.

Figure 2(f) shows the phononic power conversion efficiency as a function of phononic pump power at a fixed detection frequency of $f_3 = 626.6$ MHz for the case where the electrical contacts to the semiconductor layer are left open or shorted together, which is expected to affect the power conversion efficiency as discussed in Supplementary Note 2. A maximum phononic power conversion efficiency of (16±6)% is realized for a phononic pump power of 13 mW. The modeled phononic power conversion efficiency is also shown in Fig. 2(f). according to a model we have developed to theoretically assess the phononic frequency conversion processes. As described in detail in Supplementary Note 3, we adapt multiple models that have been previously developed for bulk piezoelectric semiconductors with plane wave propagation[28-30] to describe the nonlinear interactions we study here in the regime where $\beta l \ll 1$ where $\beta$ is the wave vector for any of the piezoelectric phonons in the system and $l$ is the charge carrier mean free path. No models exist in the literature that can capture the charge carrier relaxation and diffusion dynamics of the acoustoelectric heterostructure. To address this, we have combined plane wave[28-30] and heterostructure[31,32] models of the acoustoelectric interaction. In our model, we also account for the finite extent of the waves to capture the propagation and interaction of the phononic modes in the planar piezoelectric waveguide of the acoustoelectric heterostructure.

Agreement between the experimental results and the model is achieved, up to a pump power of approximately 10 mW, beyond which the theoretical values exceed the experimental value. The deviation of the experimental values from the theoretical values at large pump power could be due to several effects. First, as discussed below, we observe second harmonic generation and cascaded third-harmonic generation for the pump in laser vibrometry data and therefore expect the pump power available for sum frequency generation to be reduced at high powers. Second, also as discussed below, we observe strong third-order nonlinear effects at high power that we expect will lead to self-phase modulation and cross-phase modulation that will degrade the phase matching at high powers. Finally, it is also possible that at these large pump powers the input transducer is being thermally detuned from its low-power resonance frequency.

Supplementary Note 4 describes a control experiment carried out on a device where the semiconductor was removed but is otherwise identical, such that the source of the nonlinearity is from the LiNbO$_3$ alone. The peak electrical power conversion efficiency achieved is $(0.06\pm0.03)$% and the peak phononic power conversion efficiency achieved is $(0.8\pm0.5)$% at a phononic pump power of 0.01 mW. Therefore, the achieved phononic power conversion efficiency with the In$_{0.53}$Ga$_{0.47}$As semiconductor film is 20X larger, but the power-dependence of the conversion efficiency differs between the two. In Supplementary Note 4, we also show data for the power conversion efficiency as a function of frequency for the control device at a phononic pump power of 13 mW. The larger pump power does not result in larger power conversion efficiency for the LiNbO$_3$-only device as it does for the acoustoelectric heterostructure device. Ideally, we could use the two devices to directly extract the acoustoelectrically induced change in the nonlinear phonon susceptiblity responsible for the mixing. However, there are several factors that make a direct comparison of the nonlinearity challenging. First, the phase mismatch is expected to be power dependent with the In$_{0.53}$Ga$_{0.47}$As semiconductor as opposed to the LiNbO$_3$ control device and this effect is not yet included in our model. Furthermore, for the materials we utilize here, the acoustoelectric loss is large, and this loss mechanism is also expected to be power-dependent. Finally, the acoustoelectric nonlinearity itself is a complex quantity, unlike in nonlinear optics where the nonlinearity is purely real. One consequence of a complex nonlinearity is nonlinear loss, which, in combination with the previously discussed nonlinearities, further complicates the power dependence of the nonlinear mixing. With that in mind, we focus on quantification of the increase in nonlinear mixing efficiency and will investigate the changes in nonlinear susceptibility in future work. The previous state-of-the-art for three-wave mixing in LiNbO$_3$ is for a piezoelectric acoustic wave device carrying out an acoustic convolution,[22] where the electrical power conversion efficiency was $4 \times 10^{-5}$%. Therefore, here we have shown a 1500X improvement in the electrical power conversion efficiency for LiNbO$_3$ alone compared to the previous state-of-the-art, potentially due to optimization of $k^2$ and the input pump power, and an 32500X improvement with the semiconductor film heterogeneously integrated. As discussed in detail below, material optimization or alternate semiconductor materials that simultaneously decrease the acoustoelectric loss while increasing the acoustoelectric nonlinearity, combined with approaches to minimize phase mismatch, will ultimately provide the largest power conversion efficiency at the smallest pump powers for the acoustoelectric heterostructure approach for phononic three-wave mixing that we present here.

We also measured, with a similar experimental procedure, difference frequency generation (see Supplementary Note 5), which is another parametric three-wave mixing process where energy conservation requires that a phonon at $\omega_3$ is annihilated to create phonons at $\omega_2$ and $\omega_1 = \omega_3 - \omega_2$. In this case, the interdigital phonon transducers are designed such that the input pump has a frequency of $f_3 = 366.8$ MHz, the input signal has a frequency of $f_2 = 253.3$ MHz, and the output transducer has a frequency of $f_1 = f_3 - f_2 = 113.3$ MHz. As shown in Supplementary Note 5, we achieve a maximum phononic power conversion efficiency of $(1.0 \pm 0.1)\%$ with an phononic pump power of 5 mW.

The overall frequency conversion process has a strong dependence on the phase mismatch, the interaction length, the modal overlap, and the material parameters of the acoustoelectric heterostructure. We esimate the current phase mismatch for sum frequency generation to be 113 cm$^{-1}$ and for difference frequency generation to be 213 cm$^{-1}$ (see Supplementary Note 6), which significantly limits the achievable phononic power conversion efficiency for both sum and difference frequency generation. Therefore, a systematic approach to minimize or eliminate phase mismatch while optimizing the interaction length could lead to larger demonstrated power conversion efficiencies at lower pump powers compared to what has been reported here.

**Drift velocity tuning of the power conversion efficiency**
We next investigate the impact of modifying the ratio between the charge carrier drift velocity and the acoustic velocity to phononic sum frequency generation efficiency mediated by the acoustoelectric nonlinearity. Applying an external drift field to the charge carriers in an acoustoelectric heterostructure can provide large phononic small-signal gain or attenuation, depending on the size of the applied drift field and relative propagation direction of the phonons and carriers.[33-35] Supplementary Note 7 includes a plot of the theoretical linear and nonlinear acoustoelectric coefficients as a function of the ratio between the drift and acoustic velocities. At a velocity ratio exceeding 1, two effects occur simultaneously. First, the nonlinear coefficient, $\eta$ (Supplementary Note 3), increases in absolute value. Second, the amplitudes of the pump, signal, and output phonon fields are all amplified as they travel through the interaction region, providing both effective increases in the nonlinear interaction due to larger pump and signal waves *and* amplification of the output field. Therefore, it is expected that the velocity ratio can drastically tune the power conversion efficiency.

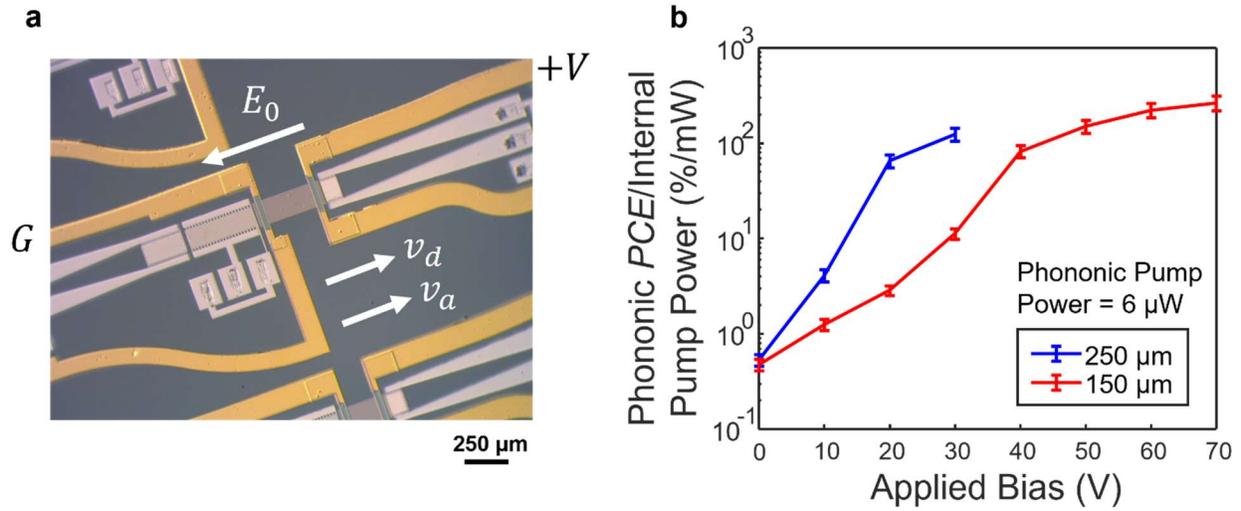

**Fig. 3: Control of the nonlinearity and mixing efficiency.** (a) Microscope image of the device indicating the positive voltage electrode ($+V$), the ground electrode ($G$), the direction of the static electric field ($E_0$), the direction of the charge carriers with a drift velocity, $v_d$, and the direction of the phonon propagation with an acoustic velocity, $v_a$. (b) Measured power-normalized power conversion efficiency ($PCE$) as a function of applied bias. The pump power was fixed at 6 µW for all applied bias values.

Experimentally, the velocity ratio is modified by applying a static electric field to the charge carrier system as shown in Fig. 3(a). Figure 3(b) shows a plot of the measured power-normalized phononic power conversion efficiency (phononic power conversion efficiency per unit phononic pump power in units of %/mW) as a function of applied bias for two different lengths of interaction region, 150 µm and 250 µm; here, the phononic pump power is fixed at 6 µW. Therefore, as shown in Fig. 3(b), for a fixed pump power, both the conversion efficiency and the power-normalized conversion efficiency increase with increasing applied bias. We find that for a 250 µm long interaction length, we achieve a maximum power-normalized conversion efficiency of $(270 \pm 50)$%/mW at an applied bias of 70 V, which corresponds to a drift field of 2.8 kV/cm. For the previous measurement results reported with no bias and variable pump power (Fig. 2f) our maximum value for the power-normalized conversion efficiency is $(1.3 \pm 0.3)$%/mW. This represents a >200x increase in power-normalized conversion efficiency for the sum-frequency conversion process at low pump powers. It should be noted that this ability to reconfigure the power conversion efficiency allows for interesting technological applications, such as gain switching of oscillators or dynamically optimizing conversion efficiency to extend the dynamic range of input signals with fixed radio frequency powers.

**Four-wave mixing**

As in nonlinear optics, we can also examine four-wave mixing processes in the system that takes advantage of $X^3_{AE}$ to see their enhancement relative to nonlinear phononic systems without acoustoelectric mediation. The possible range of interactions in four-wave mixing is large,[36] but here we focus on cascaded degenerate four-wave mixing (CDFWM) interactions where the input and output frequencies are close together. This allows us to study the third-order nonlinear process with near-perfect phase matching and to directly compare to a $LiNbO_3$ device where the $In_{0.53}Ga_{0.47}As$ semiconductor layer has been removed without the complication of variations in phase mismatch between the two cases as in three-wave mixing. LiNbO$_3$ itself is the current state-of-the-art single crystal material for microwave frequency four-wave phonon mixing.[27] We can thus directly measure and compare the efficiency of phononic four-wave mixing frequency conversion with and without the presence of the acoustoelectric nonlinearity. We follow the analysis previously developed to analyze phononic four-wave mixing in LiNbO$_3$ on sapphire channel waveguides,[27] which also provides an additional experimental result for comparison. While CDFWM in an acoustoelectric heterostructure was also recently reported,[26] they did not identify the order of nonlinearity or four-wave mixing process used, did not describe its sensitivity to momentum or phase mismatch, and did not characterize the efficiency or internal (phononic) nonlinearity, making it challenging to compare to other systems.

During the nonlinear interaction, cascaded four-wave mixing leads to the generation of equally spaced sidebands on either side of the two pumps, with the frequency spacing between sidebands set by the frequency spacing between the pumps. The broadband phase matching supports the generation of a comb of sidebands and ultimately modifies the power conversion efficiency for the first two sidebands on either side. As shown in Fig. 4(a), here we focus on two degenerate four-wave mixing processes that occur in the system simultaneously. In the first, two phonons at $\omega_1$ are annihilated to create a phonon at $\omega_2$ and $\omega_{112} = \omega_1 + \omega_1 - \omega_2$; in the second, which occurs concurrently, two phonons at $\omega_2$ are destroyed to create a phonon at $\omega_1$ and $\omega_{221} = \omega_2 + \omega_2 - \omega_1$. The phase mismatch for the first and second third-order interactions shown in Fig. 4(a) is characterized by $\Delta\beta = \beta_{112} + \beta_2 - \beta_1 - \beta_1$ and $\Delta\beta = \beta_1 + \beta_{221} - \beta_2 - \beta_2$, respectively. Figure 4(b) shows a microscope image of the device developed to study phononic four-wave mixing mediated by an acoustoelectric nonlinearity, with important experimental aspects of the measurement labeled. Additional details on the experimental setup and measurement are given in the Methods. Briefly, an interdigital phonon transducer with a resonance frequency of $f = 1.076$ GHz is utilized to simultaneously generate the pump phonons at $f_1 = 1.0755$ GHz and $f_2 = 1.0765$ GHz such that they are offset from each other by 1 MHz. These two phonons are then transmitted through a 250 μm long region of patterned In$_{0.53}$Ga$_{0.47}$As to support the acoustoelectric interaction. During the propagation, phonons at $f_{112}$ and $f_{221}$ are generated with equally spaced frequencies from the incident phonons due to the parametric four-wave mixing processes as shown in Fig. 4(a). Electrical contact is made to the In$_{0.53}$Ga$_{0.47}$As to support an open or shorted boundary condition for the acoustoelectric current (Supplementary Note 2). The phononic output, including both the incident and generated frequencies, is then converted to an electrical signal by the output interdigital transducer and is measured on a spectrum analyzer. An example output spectrum is shown in Fig. 4(c) with the detected signals at $f_1, f_2, f_{112}$, and $f_{221}$ labeled. As expected, phonons at additional frequencies are also generated due to the cascading nonlinear effect.

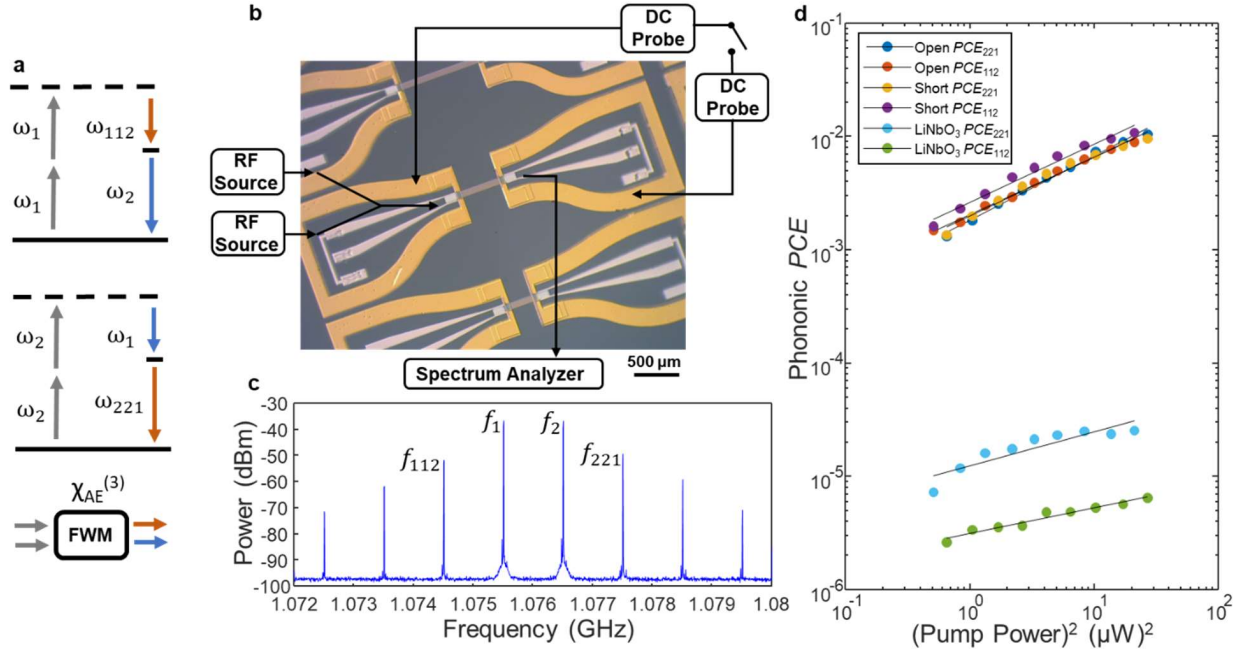

**Fig. 4: Four-wave mixing.** (a) Phonon energy level diagrams for generating $f_{112}$ and $f_{221}$ in four-wave mixing processes mediated by a third-order acoustoelectric nonlinearity, $X_{AE}^3$. (b) Microscope image of the two-port device utilized to study four-wave mixing in the acoustoelectric heterostructure. Radio frequency (RF) sources, a power combiner, and an input interdigital transducer (IDT) are utilized to generate phonons at $f_1$ and $f_2$ with a 1 MHz offset while an output IDT and spectrum analyzer are utilized to read out the spectrum after the nonlinear mixing process. (c) Measured four-wave mixing spectrum and (d) measured phononic power conversion efficiency (*PCE*) as a function of phononic pump power squared.

A plot of the phononic power conversion efficiency as a function of the square of the phononic pump power is shown in Fig. 4(d). We apply a previously developed method for phononic four-wave mixing[27] to determine the phononic power conversion efficiency values from the measured spectra, calculate the four-wave mixing coefficients from Fig. 4(d), and calculate the modal nonlinear coefficients (further details can be found in Supplementary Note 8). Table 1 summarizes the four-wave mixing coefficients and the modal nonlinear coefficients for the case of the $In_{0.53}Ga_{0.47}As$-$LiNbO_3$ acoustoelectric heterostructure and $LiNbO_3$ alone. The $LiNbO_3$ planar waveguide measurements were made using a device on the same wafer with identical geometry but without any $In_{0.53}Ga_{0.47}As$ present. Also shown in Table 1 are the results for the FWM coefficient and modal nonlinear coefficient from the literature for a $LiNbO_3$ channel waveguide.[27] As shown in Table 1, the four-wave mixing coefficients for the acoustoelectric heterostructure are two orders of magnitude larger than for the $LiNbO_3$ planar piezoelectric waveguide alone. For example, the 221 process for $In_{0.53}Ga_{0.47}As$-$LiNbO_3$ with a shorted electrical boundary condition to the $In_{0.53}Ga_{0.47}As$ has a four-wave mixing coefficient of $1700 \pm 800$ mW$^{-2}$, which is greater than 200X larger than the $LiNbO_3$ planar waveguide without the acoustoelectric nonlinearity. The significant level of error occurs here as the model's assumption that the efficiency scales as the square of the input pump power is an approximation and does not completely capture the power-dependent dynamics in the system. The FWM coefficients extracted from the experimental data can in turn be used to approximate a modal nonlinear coefficient, which quantifies the strength of the nonlinearity normalized to the input power and the effective length $L_{\text{eff}} = [1 - \exp(-\alpha L)]/\alpha$ where $L$ is the interaction length and $\alpha$ is the propagation loss. For our devices, $L = 0.25$ mm and $L_{\text{eff}} = 0.21$ mm and 0.24 mm for the structures without and without the InGaAs, respectively. The shorter effective length for the acoustoelectric heterostructure is due to the larger $\alpha$ from the acoustoelectric attenuation.[34] As shown in Table 1, our modal nonlinear coefficients with the $In_{0.53}Ga_{0.47}As$ are >10X larger when compared to the $LiNbO_3$ waveguide devices. Therefore, the third-order nonlinearity supported in our acoustoelectric heterostructure provides a significantly larger nonlinear interaction strength for piezoelectric phonons compared to the current state-of-the-art.[27] Further, it can be shown that the modal nonlinear coefficients for four-wave mixing interactions are inversely proportional to the cross-sectional area.[37] Given the ~9λ width of the waves used in this demonstration, using single-mode phononic channel waveguides is expected to increase the modal nonlinearity by a factor of nearly 20.

Table 1: Measured four-wave mixing coefficients

| Process | Device | FWM Coefficient (mW$^{-2}$) | Modal Coefficient (mW$^{-1}$mm$^{-1}$) | Study |
|---|---|---|---|---|
| 112 | $In_{0.53}Ga_{0.47}As$-$LiNbO_3$, Open | $1100 \pm 600$ | 158 | This work |
| 221 | $In_{0.53}Ga_{0.47}As$-$LiNbO_3$, Open | $1400 \pm 800$ | 178 | This work |
| 112 | $In_{0.53}Ga_{0.47}As$-$LiNbO_3$, Shorted | $1200 \pm 600$ | 165 | This work |
| 221 | $In_{0.53}Ga_{0.47}As$-$LiNbO_3$, Shorted | $1700 \pm 800$ | 196 | This work |
| 112 | $LiNbO_3$ Planar Waveguide | $2 \pm 1$ | 6 | This work |
| 221 | $LiNbO_3$ Planar Waveguide | $7 \pm 5$ | 11 | This work |
| 112 | $LiNbO_3$ Channel Waveguide | $10 \pm 5$ | 6 | Mayor et al[27] |
| 221 | $LiNbO_3$ Channel Waveguide | $20 \pm 10$ | 8 | Mayor et al[27] |

**Laser doppler vibrometry**

As shown in the section on sum and difference frequency generation, both processes are supported with similar pump and signal frequencies and similar device geometries, yet with different conversion efficiencies, which is likely due to the different degrees of phase mismatch. In addition, we observe that both second and third-order nonlinear interactions are supported in our acoustoelectric heterostructures. We further expect processes such as second and third harmonic generation of the pump and cascaded processes will occur simultaneously within these structures. While it would be possible in principle to make a series of otherwise identical structures but with different output transducers designed to collect specific mixing products, this is impractical. We instead attempt to observe the presence of all mixing components in the same device simultaneously using laser Doppler vibrometry (LDV).

The LDV measurements (Fig. 5(a)) use a scanning confocal balanced homodyne Mach-Zehnder interferometer (MZI) described in previous work.[38] We use a 1550 nm laser that reflects from the $In_{0.53}Ga_{0.47}As$ surface layer on the device. An anti-reflection-coated lensed-tapered fiber focuses the light in one arm of the MZI onto the surface of the chip. The lens produces a Gaussian mode approximately at its waist with mode-field diameter of 2 µm on the surface. Reflections from the surface are thus directly reflected back into the lensed fiber, and a magnetic circulator directs the output back into the interferometer. Surface displacements caused by the piezoelectric phonons present in the device cause phase fluctuations for the surface-reflected photons that can be converted to power fluctuations by the interferometer and measured using a spectrum analyzer to provide independent frequency resolution of the phonons. The Gaussian beam, with its 2 μm mode-field diameter, acts as a spatial filter and limits detection to frequencies of approximately 2 GHz (given the ~4000 m/s acoustic velocities).

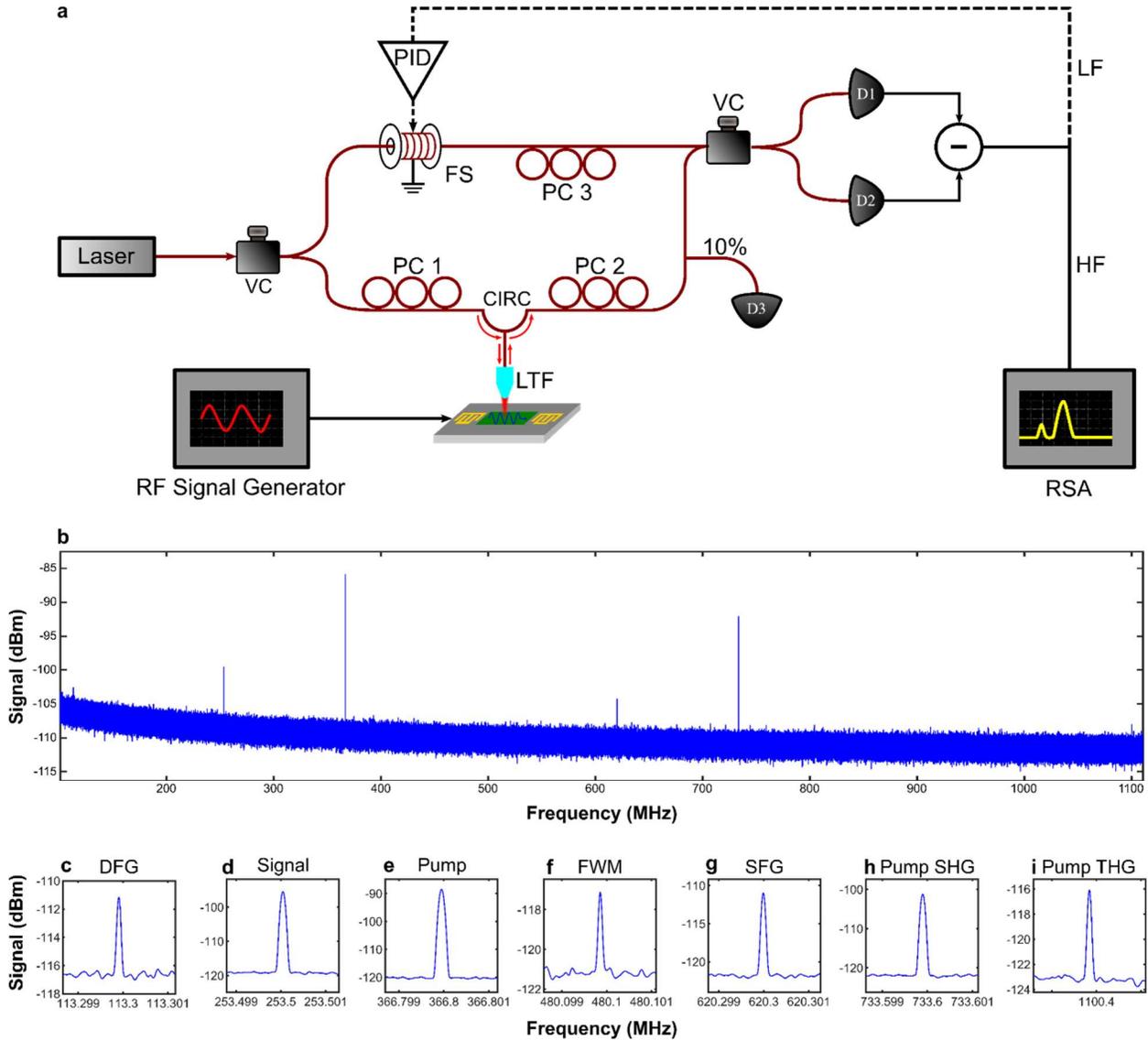

**Fig. 5: Large bandwidth frequency conversion measurements.** (a) Schematic of the laser doppler vibrometer. The laser is split into two arms utilizing a variable coupler (VC). The light in the signal arm is then sent through a polarization controller (PC), a circulator (CIRC), and is focused on the acoustoelectric heterostructure surface via a lensed tapered fiber (LTF). A 10% pickoff to a detector allows for simultaneous confocal microscopy. A PID controller actuates a fiber stretcher (FS) using feedback from the low frequency (LF) to maintain the local oscillator arm at a known phase. The high frequency (HF) output is analyzed using a real-time spectrum analyzer (RSA). (b) Interferometer signal measured using the real-time spectrum analyzer (RSA). Pump frequency 366.8 MHz driven at 22 dBm, signal frequency 253.5 MHz driven at 0 dBm. Peaks corresponding to difference frequency generation at 113.3 MHz, sum frequency

generation at 620.3 MHz, pump second harmonic at 733.6 MHz, and pump third harmonic at 1,100.4 MHz are visible. (c)-(i) Detail of peaks for difference frequency generation, signal, pump, four-wave mixing, sum frequency generation, pump second harmonic, and pump third harmonic respectively.

The device designed for difference frequency generation (Supplementary Note 5) was utilized for the measurement. Figures 5(b-i) show the captured spectra. Peaks corresponding to the pump frequency $f_p = 366.8$ MHz, signal frequency $f_s = 253.5$ MHz, difference frequency $f_d = f_p - f_s$, sum frequency $f_{sum} = f_p + f_s$, pump second harmonic $2f_p$, four-wave mixing $f_{pps} = f_p + f_p - f_s$, and pump third harmonic $3f_p$ are observed. It is important to note that many cascaded processes, such as second harmonic generation of the pump and subsequent sum-frequency generation are not distinguishable from these measurements. In future work using LDV, we will scan the position of the probe to examine both the frequency content of the mixing processes and their spatial dependence, which will allow detailed studies of their complex interactions as they propagate. This will also allow nonlinear numerical models to examine the complex mixing that occurs when so many processes can occur without excessive phase mismatch.

**Optimization of three-wave mixing nonlinear interactions and frequency conversion**
While the ability to apply bias fields to increase the effective nonlinearity is attractive, it also intrinsically amplifies the waves as they travel, which allows the excess acoustoelectric noise associated with amplification to couple into the output field, preventing quantum state preserving processes.[39] Therefore, we are highly interested in the purely parametric regime where no bias fields are applied and what the limits of frequency conversion are in that regime. In the unbiased regime, we must consider the combination of acoustoelectric loss in combination with the nonlinearity. Figure 6(a) shows a contour plot of our model's (Supplementary Note 3) predicted peak phononic power conversion efficiency for sum frequency generation as a function of the piezoelectric electromechanical coupling coefficient, $k^2$, and the phononic intensity for perfect phase matching. The phononic intensity was modified by changing the width of the interaction for a fixed input pump power of 1 mW; just as in nonlinear optics, a higher degree of transverse confinement of the waves leads to higher intensity and larger effective nonlinearities as this increases the displacement and electric fields field per unit acoustic power.

We see that maximizing both $k^2$ and the phononic intensity leads to a larger peak phononic power conversion efficiency (Fig. 6(a)) and that maximizing $k^2$ reduces the required interaction length to achieve the peak (Fig. 6(b)). At a $k^2$ of 18% and a phononic intensity of 2.5x10$^6$ W/m$^2$ (2.5 µW/µm$^2$), a peak phononic power conversion efficiency of 100% from the signal frequency to the sum frequency is theoretically achieved in an interaction length of 280 µm. The increase in phononic intensity can be achieved in our system through thinning and etching of the LiNbO$_3$ piezoelectric layer to form a channel waveguide, which could also enable phase matching through the engineering of modal dispersion in the system. A finite element method model of the z-displacement ($u \cdot \hat{z}$) for this type of structure is shown in Fig. 6(c) with a plot of the phononic intensity as a function of the waveguide width. The waveguide mode supported is a quasi-shear horizontal mode with a modeled $k^2$ value of 18.5%.

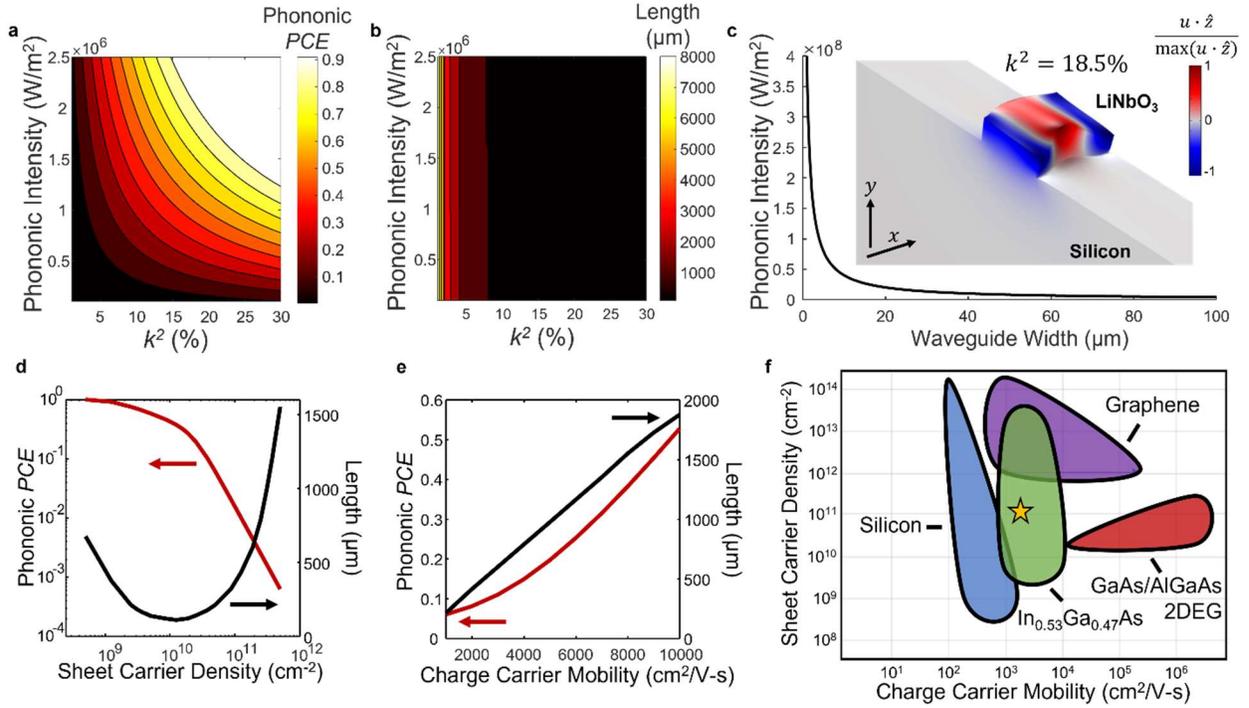

**Fig. 6: Towards stronger nonlinear interactions and larger power conversion efficiency.** (a) Peak theoretical phononic power conversion efficiency (*PCE*) as a function of $k^2$ and the phononic intensity. The mobility is fixed at 4000 cm$^2$/V-s and the charge carrier concentration is fixed at $1 \times 10^{16}$ cm$^{-3}$. (b) Interaction length to achieve the peak *PCE* as a function of $k^2$ and the intensity. (c) Phononic intensity as a function of waveguide width for a fixed pump power of 1 mW. Also shown is a finite element method model of the z-displacement, $u \cdot \hat{z}$, for a 2.5 μm thick LiNbO$_3$ channel waveguide directly on silicon supporting a quasi-shear horizontal mode with a modeled $k^2$ value of 18.5%. (d) Theoretical peak power conversion efficiency and corresponding interaction length as a function of (d) sheet carrier density and (e) charge carrier mobility. The $k^2$ value was 10% and the phononic intensity was $5 \times 10^5$ W/m$^2$. When not being varied, the sheet carrier density was fixed at $5 \times 10^{10}$ cm$^{-2}$ and the mobility was fixed at 1000 cm$^2$/V-s. (f) Semiconductor material selection chart showing material as a function of sheet carrier density and charge carrier mobility. Values for the sheet carrier densities and charge carrier mobilities are taken from relevant references[40-42] where the thickness of the semiconductor is assumed to be 50 nm for the non-2D materials.

Figures 6(d) and 6(e) show plots of the theoretical peak power conversion efficiency and the corresponding interaction length as a function of semiconductor sheet carrier density and charge carrier mobility, respectively. As can be seen, modification to the sheet carrier density can improve the phononic power conversion efficiency by orders of magnitude, and there is an optimal value for the sheet carrier density to reduce the interaction length. Figure 6(d) also suggests that the use of gate electrodes to locally modify the sheet carrier density could allow for dynamic optimization or switching of the nonlinearity. Increasing the charge carrier mobility (Fig. 6(e)) leads to a larger theoretical power conversion efficiency, but requires a longer interaction length to achieve this peak value. As our acoustoelectric heterostructure relies on bonding between the piezoelectric and semiconductor charge carrier systems, it can potentially be adapted to semiconductor materials with more favorable properties for the nonlinear frequency conversion process. A material selection chart of potential options characterizing different material classes is shown in Fig. 6(f) and includes silicon, an example of a compound semiconductor ($In_{0.53}Ga_{0.47}As$), an example of a 2D material (graphene),[40,41] and an example of a heterostructure supporting a 2D electron gas (gallium arsenide/aluminum gallium arsenide).[42] Through material selection in future work, we can potentially find the combination of sheet carrier density and charge carrier mobility to achieve nonlinear interactions in an acoustoelectric heterostructure with exceptionally large interaction strengths even beyond what is demonstrated here.

**Outlook**
The ability to generate efficient phononic frequency conversion presents an enormous open canvas for fundamental research and technological applications. As an example of materials science research that could be explored, phonons significantly contribute to heat transport properties in solids, and the ability to modify phononic nonlinearities by modifying the semiconductor properties could allow detailed studies of the specific nonlinearities that lead to processes such as thermal relaxation. This could also allow the creation of synthetic materials that have thermal properties not found in nature, such as materials with completely depleted phonon frequency regions or phonon populations that can only flow in one direction, as these momentum-preserving frequency conversion processes are highly nonreciprocal, such has been demonstrated in the photonic domain through nonlinear optical processes.[43]

The technological implications are profound. For example, three-wave mixing up-conversion and down-conversion processes are the backbone of radio-frequency signal processing, as they allow frequency conversion between radio bands and the encoding and decoding of data on carrier waves. Given recent demonstrations of acoustic amplification in the same system, one can now quite readily imagine entire radio-frequency signal processors occurring on the surface of a lithium niobate microchip or other piezoelectric substrate, greatly reducing the number of chips necessary for wireless technologies. Strong cascaded nonlinearities could be used to down-convert very high frequency radio waves where acoustic and microwave filter bandwidths are very large to lower frequencies where the filter are extremely narrow-band, and then the reverse process could allow up-conversion back to the original carrier frequency.

In the quantum realm, three-wave mixing processes such as sum-frequency generation are known to be quantum state preserving,[39] and this could provide a way to connect different microwave frequency qubits together on chips or up-convert phonons carrying quantum information to frequencies where they are more easily detectable or transducable to photons.[44,45] Another important three-wave mixing process is degenerate parametric amplification, which could allow the creation of squeezed phononic vacuum states or quantum-limited amplification of phononic quantum information, much like a Josephson parametric amplifier does for microwave photons. Finally, these giant four-wave mixing nonlinearities can be used to create self-Kerr modulation in acoustic cavities that may enable the creation of new kinds of qubits, which would then have access to the parametric amplification and state-preserving frequency conversion already discussed.

In conclusion, we have demonstrated efficient phononic three-wave and four-wave mixing nonlinearities in a thin lithium niobate film by heterogeneously integrating a high-mobility semiconductor that allows the phononic nonlinearity to be mediated by electronic nonlinearities in the semiconductor. We demonstrated three-wave mixing processes at microwave frequencies, including a $(16\pm6)$% phononic power conversion efficiency for sum-frequency generation and $(1.0\pm0.1)$% phononic power conversion efficiency for difference-frequency generation, as well as the most efficient degenerate four-wave phononic mixing to date. We presented a theoretical model that accurately predicts the sum-frequency and difference-frequency generation processes and showed experimentally that the frequency conversion can be further enhanced by the application of bias fields. Laser Doppler vibrometry was utilized to examine many three-wave and four-wave mixing processes simultaneously in the same device. Finally, we used the theoretical model to show that these nonlinearities can be greatly enhanced by confining phonons to smaller dimensions in waveguide circuits and optimizing semiconductor material properties or using 2D semiconductor materials.

**Methods**
**Acoustoelectric heterostructure and device fabrication**
The heterostructure is comprised of a 4 inch lithium niobate ($LiNbO_3$) on silicon wafer, with a 5 µm $LiNbO_3$ film thickness, and a 2 inch indium phosphide (InP) wafer. The InP wafer has a lattice-matched epitaxial stack of multiple layers of $In_{0.53}Ga_{0.47}As$ and InP. The epitaxial semiconductor stack is grown by metal-organic chemical vapor deposition (MOCVD) and consists of the following layers: 500 nm InP non-intentionally doped (NID) buffer, 3000 nm NID $In_{0.53}Ga_{0.47}As$ etch stop, 100 nm InP etch stop doped with silicon at $1\times10^{18}$ cm$^{-3}$, 100 nm $In_{0.53}Ga_{0.47}As$ contact layer doped with silicon at $2\times10^{19}$ cm$^{-3}$, 30 nm thick InP etch stop doped with silicon at $1\times10^{18}$ cm$^{-3}$, a 50 nm thick $In_{0.53}Ga_{0.47}As$ device layer, and a 5 nm NID InP capping layer. The $LiNbO_3$-silicon and $In_{0.53}Ga_{0.47}As$/InP wafers are then bonded together, with the InP capping layer in contact with the $LiNbO_3$ surface. The bond is manually initiated followed by annealing in vacuum at 100˚C. After wafer bonding, the InP substrate and buffer layer are removed in hydrochloric acid followed by removal of the subsequent $In_{0.53}Ga_{0.47}As$ etch stop layer in a 1:1:10 solution of sulfuric acid, hydrogen peroxide, and/water. The InP etch stop layer is then removed in a 1:3 mixture of hydrochloric acid, phosphoric acid, stopping on the $In_{0.53}Ga_{0.47}As$ contact layer. This highly doped $In_{0.53}Ga_{0.47}As$ layer provides an intermediary $In_{0.53}Ga_{0.47}As$ layer from the $In_{0.53}Ga_{0.47}As$ device layer to the metal electrodes to form quasi-Ohmic electrical contact to the $In_{0.53}Ga_{0.47}As$ device layer. This $In_{0.53}Ga_{0.47}As$ is patterned and then etched in a 1:1:10 mixture of sulfuric acid, hydrogen peroxide, anddeionized water followed by immediately etching the subsequent InP etch stop layer in a 1:3 mixture of hydrochloric acid/phosphoric acid. The $In_{0.53}Ga_{0.47}As$ device layer is then patterned and etched in a 1:1:10 mixture of sulfuric acid, hydrogen peroxide, and water, landing on the $LiNbO_3$ surface. A metal liftoff step, with a metal stack of 10 nm chrome/100 nm aluminum is then carried out to form the interdigital phonon transducers on the $LiNbO_3$. Electrical contact is made to the $In_{0.53}Ga_{0.47}As$ through a metal liftoff, with a metal stack of 10 nm Ti, 500 nm Au, 500 nm Ag, and 100 nm Au, over the patterned $In_{0.53}Ga_{0.47}As$ contact layer. The semiconductor properties of the 50 nm thick $In_{0.53}Ga_{0.47}As$ device layer were determined by a Hall mobility measurement using a Bio-Rad fixed-magnetic-field Hall effect measurement system. The Hall structures were patterned on the same wafer and undergo the same fabrication process flow. The average values for the Hall coefficient and Hall mobility were $-(3.5\pm0.1)\times10^3$ m$^2$C$^{-1}$ and $4{,}220\pm40$ cm$^2$ V$^{-1}$s$^{-1}$, respectively, as taken from measurements on two separate structures. Combined with the measured sheet resistance from the Hall structures, this gives a doping concentration of approximately $3 \times 10^{16}$ cm$^{-3}$ for the In$_{0.53}$Ga$_{0.47}$As device layer. The two Hall structure measurements were taken in similar locations on the wafer while there is an expected cross-wafer variation of ~12% in sheet resistance across the wafer primarily due to variations in carrier concentration due to non-uniformity in disilane delivery during the epitaxial growth.

**Reflection measurements**

Interdigital transducer reflection measurements were carried out on a network analyzer on a custom radio frequency probe station with ground-signal-ground radio frequency probes\. A full one-port calibration (short, load, open) was carried out with an impedance standard substate before all measurements to calibrate out the response of the cables and probes. For all measurements, the radio frequency power on the network analyzer was set to -10 dBm and the reflection coefficient ($S_{11}$) was measured as a function of frequency to characterize each interdigital transducer.

**Sum and difference frequency generation power conversion efficiency measurements**

Sum and difference frequency power conversion efficiency measurements were made on a custom radio frequency-DC probe station with three ground-signal-ground radio frequency probes and two DC probes. To generate the pump phonons, a radio frequency signal generator is set to output at a fixed frequency, equal to the resonant frequency of the interdigital transducer for the pump. This output then goes through a radio frequency amplifier before being sent through the interdigital transducer to generate the pump phonons. A network analyzer is utilized to source a frequency resonant with the signal interdigital transducer while detecting at the sum or difference frequency. The transmission measurement ($S_{21}$) on the network analyzer then directly measures the electrical power conversion efficiency from the signal frequency to the sum or difference frequency. The 10 MHz internal reference of the radio frequency signal generator was connected to the network analyzer to phase-lock the network analyzer to the signal generator. A scalar-mixer calibration was carried out before all measurements to calibrate out the response of the cables and probes for the measured $S_{21}$ values. For each data point, 1601 consecutive measurements (corresponding to the maximum number of points that can be taken in a single trace on the network analyzer) were taken. The phononic power conversion efficiency values then are based on the average of these 1601 points along with the measured transducer conversion losses. The error in the reported phononic power conversion efficiencies comes from two sources:one is the standard deviation of the 1601 measurements and the second is the inferred error in the interdigital transducer conversion losses.

**Four wave mixing measurements**

To carry out the four wave mixing measurements, we utilized two radio frequency signal generators and a spectrum analyzer. The outputs of the signal generators go to a power combiner and then to the single interdigital input transducer of the device to generate the pump phonons. The output from the device goes to the spectrum analyzer, To determine the phononic input pump power, a radio frequency power meter was used to measure the loss in the cable and acoustic delay line test structures were utilized to determine the losses incurred during the electromechanical conversion process.

**Acknowledgements**

Supported by the Laboratory Directed Research and Development program at Sandia National Laboratories, a multimission laboratory managed and operated by National Technology and

**Data availability**
The data that support the findings of this study are available from the corresponding author upon reasonable request.

**Conflict of interest**
The authors declare that they have no conflict of interest.

**Author contributions**
L.H. and M.E. came up with the device concepts and experimental implementations. M.E. developed the laser doppler vibrometry system. M.K. performed measurements with the laser doppler vibrometer with input from L.H., N.O., and M.E.. L.H., B.S., M.M., S.W., S.A., T.A.F. and M.E. designed the devices and fabrication process flow. M.M., S.W., B.S., and S.A. fabricated the devices. L.H., M.K., B.S., and S.S. performed the measurements. L.H. and M.E. carried out all modeling. L.H., M.K., and M.E. analyzed all data with input from N.O. The manuscript was written by L.H., M.K., and M.E. and all authors have given approval of the final version.

**Supplementary Note 1**
**Acoustic delay line test structures**
To determine the internal pump power and internal power conversion efficiency, we utilized acoustic delay line test structures in the lithium niobate (LiNbO$_3$)-silicon. Eight different types of acoustic delay line test structures were utilized: acoustic wavelength of 6.4 μm and aperture of 100 μm, acoustic wavelength of 11 μm and aperture of 100 μm, acoustic wavelength of 16 μm and aperture of 100 μm, acoustic wavelength of 11 μm and aperture of 240 μm, acoustic wavelength of 16 μm and aperture of 240 μm, acoustic wavelength of 39 μm and aperture of 240 μm, acoustic wavelength of 11 μm and aperture of 100 μm with a signal interdigital transducer with a wavelength of 16 μm and aperture of 100 μm placed in the acoustic path, and acoustic wavelength of 11 μm and aperture of 240 μm with a signal interdigital transducer with a wavelength of 16 μm and aperture of 240 μm placed in the acoustic path. For each of the eight different types of acoustic delay line test structures, acoustic delay lines of varying separation lengths between the interdigital transducers utilized to generate and detect acoustic waves were fabricated and characterized. From the loss as a function of length, we can extract the propagation loss. The remaining losses are assumed to be from the transducers. Therefore, by splitting the remaining losses in half we can determine the loss for a single interdigital transducer. For the pump (11 μm acoustic wavelength) test structures with the signal transducer in the acoustic path, we measured the increase in loss due to the signal transducer. Supplementary Fig. 1 shows data for a single type of acoustic delay line test structure and Supplementary Table 1 summarizes the losses for all test structures.

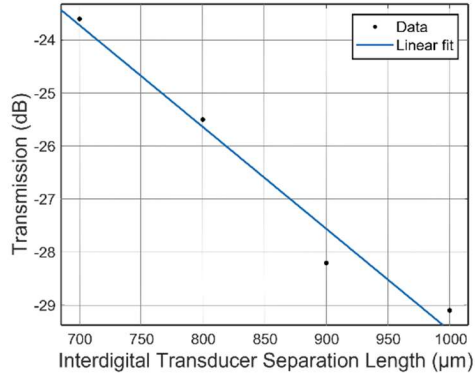

**Supplementary Fig. 1: Acoustic delay line test structures.** The measured transmission as a function of separation length between two interdigital transducers utilized to generate and detect acoustic waves is shown. For this test structure, the transducer is designed to be resonant at approximately 255 MHz, corresponding to an acoustic wavelength of 16 µm, and the aperture of the interdigital transducer is 100 µm.

Supplementary Table 1: Loss per interdigital transducer

| Wavelength (µm) | Aperture (µm) | Loss (dB) | Note |
|---|---|---|---|
| 11 | 100 | $7 \pm 1$ | |
| 16 | 100 | $6 \pm 3$ | |
| 6.4 | 100 | $5 \pm 1$ | |
| 11 | 240 | $6 \pm 2$ | |
| 16 | 240 | $6 \pm 1$ | |
| 39 | 240 | $7 \pm 1$ | |
| 11 | 100 | $12 \pm 2$ | With signal transducer in path |
| 11 | 240 | $11 \pm 3$ | With signal transducer in path |

**Supplementary Note 2**
**Electrical boundary condition to semiconductor**
In the experiments in this work, the semiconductor boundary conditions could be modified by the application of electrical boundary conditions to the electrodes at the ends of the semiconductor. Though rigorous analysis must be done in the future to thoroughly understand the impact these electrical boundary conditions make, we at least attempt here to explain why they are expected to impact the performance of the mixer devices. In the case that the two contacts are shorted together, the electrical current generated by the acoustoelectric interactions at one end of the device must be equal to the current at the other end of the device, especially considering the very large resistance of the semiconductor itself. In the case that the contacts are allowed to float or be an open circuit, then the currents at both ends of the device must go to zero. While the AC currents generated by the acoustic waves and their nonlinear interactions can do this for fortuitous values of the electronic current's phase, the DC currents produced by the acoustic waves[31] cannot. We thus presume that the DC electric currents must satisfy these conditions by setting up counter-propagating DC currents, and these currents, appearing to be counterpropagating drift currents from the perspective of the waves that generate them, would be expected to attenuate the acoustic wave. Thus, we generically expect to have higher mixing efficiency for the case of shorted electrodes. While this is certainly true for our best values of sum-frequency generation, as can be seen in Fig. 2(f) in the main text, it can also be seen that the general relationship between the mixing efficiency and the electrical boundary condition is not that simple. In this work, we measured efficiencies with both boundary conditions, but solving the nonlinear coupled mode equations in these systems in the presence of realistic boundary conditions is something that should be done in future work to understand the impact on the mixing efficiencies.

**Supplementary Note 3**
**Model development: piezoelectric acoustic wave nonlinear mixing when coupled to a semiconductor charge carrier system**

Here we develop a theoretical model for nonlinear phononic three-wave mixing interactions induced by an acoustoelectric nonlinearity in a heterostructure comprised of an acoustically thin semiconductor film and a piezoelectric acoustic wave film. Previous models have been developed of acoustoelectrically mediated phonon interactions in an infinite "isotropic" piezoelectric semiconductor,[28-30] but these models differ in many important ways that are necessary to explain our experimental results. First, the carrier dynamics inside a thin, non-piezoelectric semiconductor are considerably different from those in a bulk piezoelectric semiconductor, including significant differences between carrier relaxation dynamics due to drift, dissipation, and diffusion. Second, the response to the acoustic wave depends on the spatial dependence of the electric field, whereas in any previously developed model[28-30] the plane-wave nature of the acoustic, electric, and electronic waves are independent of any spatial variations in those fields. Moreover, because of the separation of the two media, it is in fact the evanescent electric field of the piezoelectric phonon and the spatial dependence of that field that results in the coupling. Third, existing models cannot readily take into account the enhancement of the nonlinearities due to confinement of the acoustic wave, which increases the electric field for a given phonon density (number of phonons/length) in the waveguide.

The structure of our planar phononic waveguide consists of a $LiNbO_3$ guiding material and a silicon substrate where the acoustic velocity in the silicon exceeds the acoustic velocity for the acoustic modes supported in the $LiNbO_3$. The eigenmodes of the waveguide can be found by solving the coupled acoustic and electromagnetic wave equations for the system and a complex reciprocity relation, which allows modal decomposition of arbitrary fields in terms of power-orthogonal waveguide modes.[46] These waveguide modes have normalized transverse field patterns that are assumed to be constant along the propagation direction ($x$) with characteristics that depend on the distribution of acoustic and electromechanical material properties along the waveguide cross-sectional profile ($y$).

In the case of a piezoelectric semiconductor, the coupled-wave equations for the displacement amplitude, $\hat{u}_m$, under the assumption that the change in displacement is small with respect to the acoustic wavelength, are given by

$$\frac{d\hat{u}_1}{dx} = -\hat{\alpha}_1 \hat{u}_1 + \hat{\eta}_1 \, \hat{u}_2^* \hat{u}_3 e^{i\Delta\beta x} \tag{S1a}$$

$$\frac{d\hat{u}_2}{dx} = -\hat{\alpha}_2 \hat{u}_2 + \hat{\eta}_2 \, \hat{u}_1^* \hat{u}_3 e^{i\Delta\beta x} \tag{S1b}$$

$$\frac{d\hat{u}_3}{dx} = -\hat{\alpha}_3 \hat{u}_3 + \hat{\eta}_3 \, \hat{u}_1 \hat{u}_2 e^{i\Delta\beta x} \tag{S1c}$$

where $\Delta\beta = \beta_1 + \beta_2 - \beta_3$ is the wave vector mismatch, $\beta_m$ are the wave vectors, $\hat{\eta}_m$ are the nonlinear coefficients given by

$$\hat{\eta}_1 = i\frac{e^3\mu}{4c\varepsilon^2 v_a}\frac{\omega_c}{\omega_1}\frac{2\gamma + i\omega_1/\omega_D}{\Gamma_1\Gamma_2^*\Gamma_3}\beta_2\beta_3 \tag{S2a}$$

$$\hat{\eta}_2 = i\frac{e^3\mu}{4c\varepsilon^2 v_a}\frac{\omega_c}{\omega_2}\frac{2\gamma + i\omega_2/\omega_D}{\Gamma_1^*\Gamma_2\Gamma_3}\beta_1\beta_3 \tag{S2b}$$

$$\hat{\eta}_3 = i\frac{e^3\mu}{4c\varepsilon^2 v_a}\frac{\omega_c}{\omega_3}\frac{2\gamma + i\omega_3/\omega_D}{\Gamma_1\Gamma_2\Gamma_3}\beta_1\beta_2 \tag{S2c}$$

and $\hat{\alpha}_m$ are the acoustoelectric linear attenuation/gain coefficients given by

$$\hat{\alpha}_m = \frac{e^2}{2c\varepsilon}\frac{\omega_c}{v_a\gamma}\left[1 + \frac{\omega_c^2}{\gamma^2\omega_m^2}\left(1 + \frac{\omega_m^2}{\omega_c\omega_D}\right)^2\right]^{-1} \tag{S3}$$

where $e$ is the piezoelectric coupling constant, $\mu$ is the charge carrier mobility, $c$ is the elastic constant, $\varepsilon$ is the material permittivity, $v_a$ is the acoustic velocity, $\omega_m$ is the acoustic radial frequency, $\omega_c = \sigma/\varepsilon$ is the dielectric relaxation frequency, $\sigma$ is the conductivity, $\omega_D = v_a^2/D_n$ is the diffusion frequency, $D_n = \mu k_B T/q$ is the diffusion coefficient, $k_B$ is Boltzmann's constant, $T$ is the temperature, $q$ is the elementary charge, $\gamma$ is a term that denotes the ratio between the acoustic velocity, $v_a$, and the charge carrier drift velocity, $v_d$, according to $\gamma = 1 - v_d/v_a$, and $\Gamma_m = \gamma + i(\omega_c/\omega_m + \omega_m/\omega_D)$.[28] The term $e^2/c\varepsilon$ is approximately equal to the volume-wave electromechanical coupling coefficient, $K^2$, which quantifies the piezoelectric strength of the material.[47]

We first need to adapt the expressions for the nonlinear (Eqn. S2) and linear (Eqn. S3) acoustoelectric coefficients from a `` to the acoustoelectric heterostructure studied in this work. To do this, we begin with Eqn. S3 as there exists an expression for the linear attenuation/gain coefficient in an acoustoelectric heterostructure, $\alpha_m$, that has been derived via perturbation theory.[32] This allows us to compare our modifications to the expression for the linear acoustoelectric coefficient (Eqn. S3) to the result derived by perturbation theory for an acoustoelectric heterostructure. If suitable agreement between the two approaches is achieved, then we can also apply our modifications to the nonlinear acoustoelectric coefficients (Eqn. S2) to parameterize the nonlinear interaction in the heterostructure.

For the case of an acoustoelectric heterostructure, the expression for the linear acoustoelectric attenuation/gain coefficient, as determined by a perturbative approach, is given by

$$\alpha_m = \frac{1}{2}\frac{(v_d/v_a - 1)\omega_c\varepsilon_s w Z_{a,m}(h)\beta\tanh(\beta d)}{(v_d/v_a - 1)^2 + (R\omega_c/\omega_m + H)^2} \tag{S4}$$

where $Z_{a,m}(h) = 2|\Delta v(0)|e^{-2\beta h}/wv_a\omega_m(\varepsilon_0 + \varepsilon_p)$ is the interaction impedance, $\Delta v(0)$ is the perturbation in velocity of the piezoelectric acoustic wave when the surface electrical boundary condition of the piezoelectric is changed from electrically open to shorted, $h = g e_G$ where $g$ and $e_G$ are the thickness and relative permittivity, respectively, for the gap between the piezoelectric and semiconducting materials in the heterostructure, $w$ is the width of the interaction region, $\omega_c = \sigma/\varepsilon_s$ is the dielectric relaxation frequency, $\varepsilon_0$ is the vacuum permittivity, $\varepsilon_p$ is the piezoelectric material permittivity, $\varepsilon_s$ is the semiconductor material permittivity, $d$ is the semiconductor thickness, $R = (\varepsilon_s/\varepsilon_0)M(\beta h)\tanh(\beta d)$ is the space-charge reduction factor, which accounts for the finite thickness of the semiconductor, $M(\beta h) = \frac{\varepsilon_0 + \varepsilon_p \tanh(\beta h)}{(\varepsilon_0 + \varepsilon_p)(1 + \tan(\beta h))}$, and $H = \sqrt{\frac{\omega_c}{\omega_D} \frac{\tanh(\beta d)}{\tanh\left((\sqrt{\omega_c \omega_D}/v_a)d\right)}}$ is a term that arises from semiconductor charge carrier diffusion.[32]

The expression for (S3), adapted for our acoustoelectric heterostructure, is given by

$$\alpha_m = \frac{k_m^2}{2} \frac{\omega_{c,m}}{v_{a,m} \gamma_m} \left[ 1 + \frac{\omega_{c,m}^2}{\gamma_m^2 \omega_m^2} \left( 1 + \frac{\omega_m^2}{\omega_{c,m} \omega_D} \right)^2 \right]^{-1} \tag{S5}$$

where $k_m^2$ is the modal electromechanical coupling coefficient, which has a standard definition of $k_m^2 = 2|\Delta v_{a,m}(0)|/v_{a,m}$, which for an anisotropic material like LiNbO$_3$ depends on the material cut and the propagation direction of the piezoelectric phonon, and $\omega_{c,m} = \sigma \beta_m d / (\varepsilon_p + \varepsilon_0)$.[31] The variation in all parameters with the acoustic frequency is due to dispersion from the LiNbO$_3$ and indium gallium arsenide (In$_{0.53}$Ga$_{0.47}$As) films. The value for $\alpha$ as a function of the charge carrier mobility, $\mu$, and the charge carrier concentration, $N_d$, is plotted in Supplementary Fig. 2 for the expressions given in Eqn. S4 and Eqn. S5 with and without the inclusion of charge carrier diffusion. While we do find a difference in value of $N_d$ that maximizes $\alpha$ between the two theoretical approaches, our model correctly captures the trends in how $\alpha$ varies with the semiconductor parameters and gives values for $\alpha$ that agree well with the perturbative approach. We find that $\alpha$ is expected to reach a minimum either when $N_d$ is very low ($< 10^{14}$) or when $N_d$ is relatively large ($> 10^{16}$). For the combination of $\mu$ and $N_d$ utilized in this work, which are $\mu = 4220$ cm$^2$/V-s and $N_d = 3 \times 10^{16}$ cm$^{-3}$ (see Methods in main article), the value is 5 cm$^{-1}$ from the model developed here and 3 cm$^{-1}$ from the perturbative approach.[32] Therefore, we find that sufficient agreement is achieved.

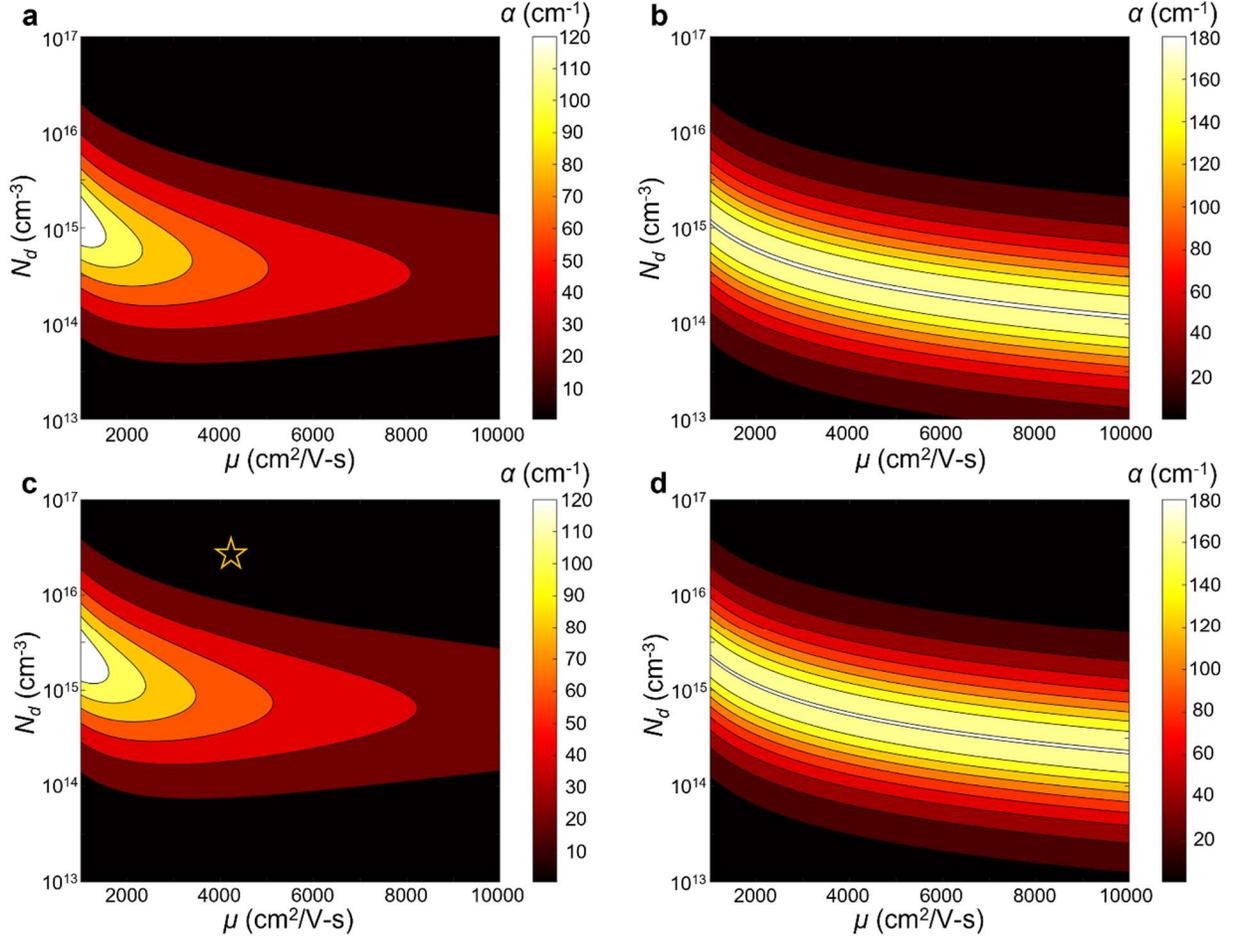

Supplementary Fig. 2: Theoretical linear acoustoelectric attenuation coefficient, $\alpha$, as a function of semiconductor mobility, $\mu$, and charge carrier concentration, $N_d$, with an expression derived from a perturbative approach (Eqn. S4)[32] (a) with and (b) without the effects of charge carrier diffusion. Plots of $\alpha$ as a function of $\mu$ and $N_d$ from our phenomenological model (Eqn. S5) (c) with and (d) without charge carrier diffusivity effects. The star marker in (c) indicates the expected value of $\alpha$ in this work given the measured values of $\mu$ and $N_d$ for the semiconductor (see Methods in main article).

We then find it suitable to apply the modifications for $k_m^2$, $\omega_{c,m}$, and dispersion to the nonlinear acoustoelectric coefficients (Eqn. S2) to obtain the following phenomenological expressions for the nonlinear acoustoelectric coefficients in the acoustoelectric heterostructure, $\eta_m$,

$$\eta_1 = i \frac{(k_1^2)^{3/2} \mu}{4 v_{a,1}} \sqrt{\frac{c_1}{\varepsilon}} \frac{\omega_{c,1}}{\omega_1} \frac{2\gamma_1 + i\,\omega_1/\omega_D}{\Gamma_1 \Gamma_2^* \Gamma_3} \beta_2 \beta_3 \tag{S6a}$$

$$\eta_2 = i \frac{(k_2^2)^{3/2} \mu}{4 v_{a,2}} \sqrt{\frac{c_2}{\varepsilon}} \frac{\omega_{c,2}}{\omega_2} \frac{2\gamma_2 + i\,\omega_2/\omega_D}{\Gamma_1^* \Gamma_2 \Gamma_3} \beta_1 \beta_3 \tag{S6b}$$

$$\eta_3 = i \frac{(k_3^2)^{3/2} \mu}{4 v_{a,3}} \sqrt{\frac{c_3}{\varepsilon}} \frac{\omega_{c,3}}{\omega_3} \frac{2\gamma_3 + i\,\omega_3/\omega_D}{\Gamma_1 \Gamma_2 \Gamma_3} \beta_1 \beta_2. \tag{S6c}$$

A plot of $|\eta_3 u_1|$ as a function of the charge carrier mobility, $\mu$, and charge carrier concentration, $N_d$, with and without charge carrier diffusion is shown in Supplementary Figs. 3(a) and 3(b), respectively. Here $u_1$ is the pump displacement, which we determine according to $|u_1| = \sqrt{P_1/(\omega_1^2 \rho v_{a,1} A_1)}$ where $P_1$ is the pump power, $\rho$ is the density, and $A_1$ is the cross-sectional area for the acoustic power flow, which is determined by the device width and a finite element method model of the acoustic mode in the waveguide cross-section. For the plots in Supplementary Fig. 3, we assume a $P_1$ of 1 mW and a device width of 100 μm. We find that the value of $|\eta_3 u_1|$ is maximized at an optimal value for $N_d$ that is relatively low (~$10^{15}$ cm$^{-3}$ for the case of no charge carrier diffusion) combined with a $\mu > 2000$ cm$^2$/V-s. For the semiconductor material parameters utilized in this work, which corresponds to $N_d = 3 \times 10^{16}$ cm$^{-3}$ and $\mu = 4220$ cm$^2$/V-s, we find that $|\eta_3 u_1| = 25$ cm$^{-1}$. Optimization of the semiconductor material parameters could lead to exceptionally larger $\eta_3$ in future work. For a mobility of $\mu = 7000$ cm$^2$/V-s and $N_d = 3 \times 10^{14}$ cm$^{-3}$, we can increase the value for $|\eta_3 u_1|$ >200X while maintaining a relatively low linear acoustoelectric attenuation (Supplementary Fig. 2c).

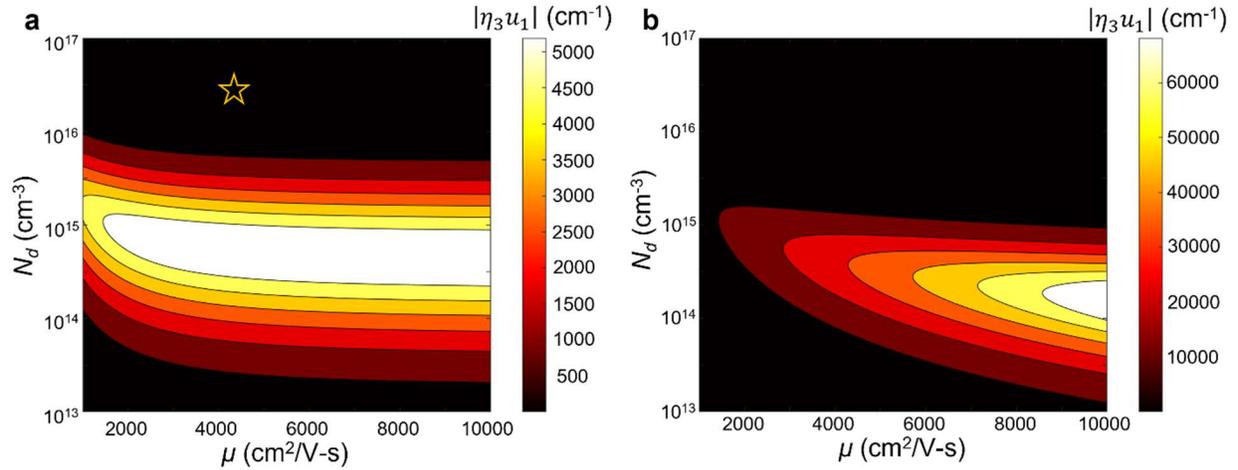

Supplementary Fig. 3: (a) Theoretical value for $|\eta_3 u_1|$ as a function of semiconductor charge carrier mobility, $\mu$, and charge carrier density, $N_d$, with and (b) without charge carrier diffusion effects. The star marker in (a) indicates the expected value of $|\eta_3 u_1|$ in this work given the measured values of $\mu$ and $N_d$ for the semiconductor (see Methods in main article).

The coupled-wave equations (Eqn. S1) are formulated by the assumption that the total fields of the piezoelectric acoustic waves being coupled are expressed utilizing the plane wave approximation. However, the planar waveguide interaction geometry utilized here makes this a guided-wave system supporting waveguide modes according to mode orthogonality as discussed above. The approach of nonlinear coupled-mode theory then applies to analyze nonlinear interactions in the phononic waveguides. We assume that the total displacement of the interacting waves, $u$, is expressed as

$$u(r,t) = \sum_m u_m(r)\exp(-i\omega_m t) \quad (S7)$$

Where $u_m(r)$ is the spatially dependent total displacement field amplitude for frequency $\omega_m$, which can be expanded in in terms of the waveguide modes, defined by an index $v$, such that

$$u_m(r) = \sum_v A_{m,v}(x) U_{m,v}(y)\exp(i\beta_m x) \quad (S8)$$

where $A_{m,v}(x)$ and $U_{m,v}(y)$ are the mode amplitudes and the normalized displacement mode fields, respectively. Here we are assuming that only one mode for each frequency is involved in the nonlinear interaction, therefore we have

$$u_m(r) = A_m(x)U_m(y)\exp(i\beta_m x). \tag{S9}$$

To adapt Eqn. S1 to the planar waveguide nonlinear interaction geometry of the acoustoelectric heterostructure, we have the following set of equations, which completes the phenomenological model of the nonlinear interaction in the acoustoelectric heterostructure:

$$\frac{dA_1}{dx} = -\alpha_1 A_1 + \kappa_1 A_2^* A_3 e^{i\Delta\beta x} \tag{S10a}$$

$$\frac{dA_2}{dx} = -\alpha_2 A_2 + \kappa_2 A_1^* A_3 e^{i\Delta\beta x} \tag{S10b}$$

$$\frac{dA_3}{dx} = -\alpha_3 A_3 + \kappa_3 A_1 A_2 e^{-i\Delta\beta x} \tag{S10c}$$

where $\alpha_m$ are the linear acoustoelectric coefficients (Eqn. S7) and $\kappa_m$ are the nonlinear coupling coefficients given by

$$\kappa_1 = \eta_1 \int_{-\infty}^{\infty} U_1^* U_2^* U_3 dy \tag{S11a}$$

$$\kappa_2 = \eta_2 \int_{-\infty}^{\infty} U_1^* U_2^* U_3 dy \tag{S11b}$$

$$\kappa_3 = \eta_3 \int_{-\infty}^{\infty} U_1 U_2 U_3^* dy \tag{S11c}$$

and $\eta_m$ are the nonlinear acoustoelectric coefficients $\eta_m$ (Eqn. S6), which exist outside of the integral as they are spatially constant values that quantify the size of the nonlinearity that arises via electron-phonon coupling in our acoustoelectric heterostructure.

**Supplementary Note 4**
**Sum frequency generation in lithium niobate**
In this supplementary note, we show experimental results taken for sum frequency generation in an identical device to that presented in Fig. 2 in the main text, except that the $In_{0.53}Ga_{0.47}As$ semiconductor film is removed. Supplementary Fig. 4(a) shows plots of both the phononic power conversion efficiency and the electrical power conversion efficiency as a function of pump power. A peak phononic power conversion efficiency of $(0.8\pm0.5)\%$ and a peak electrical power conversion efficiency of $(0.06\pm0.03)\%$ is achieved at an internal pump power of 0.01 mW. Unlike the acoustoelectric heterostructure devices, improved power conversion efficiency is not achieved at larger power powers, as shown in the plot of electrical power conversion efficiency as a function of detection frequency for an internal pump power of 13 mW (Supplementary Fig. 4(b)). As discussed in the main text, the power dependence and saturation behavior for both the nonlinearity in $LiNbO_3$ alone and the acoustoelectric nonlinearity require further study.

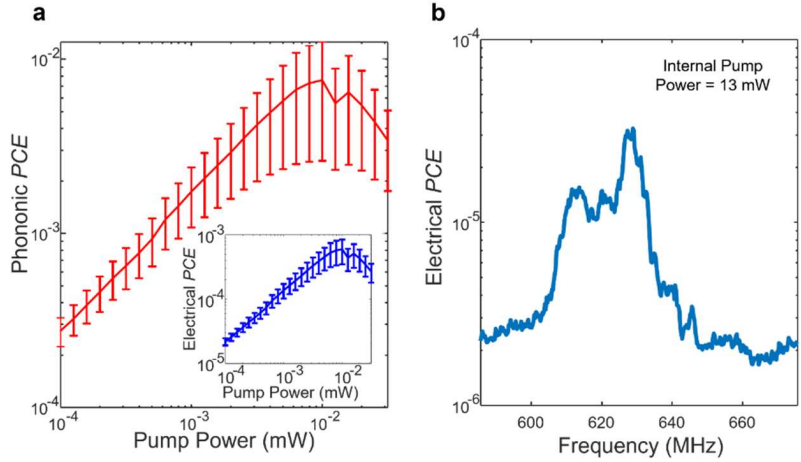

Supplementary Fig. 4: (a) Phononic power conversion efficiency (*PCE*) and electrical *PCE* (inset) as a function of internal pump power. (b) Electrical *PCE* as a function of frequency at an internal pump power of 13 mW.

## Supplementary Note 5
**Difference frequency generation**
As shown in Supplementary Fig. 5(a), difference frequency generation is another parametric three-wave mixing process where, in this case, energy conservation requires that a phonon at $\omega_3$ is annihilated to create phonons at $\omega_2$ and $\omega_1 = \omega_3 - \omega_2$. A microscope image of the developed device in the semiconductor piezoelectric system to study this frequency conversion process is shown in Supplementary Fig. 5(b). The transducer for detection is designed to have a resonance at the difference frequency $f_1 = \omega_1/2\pi = \omega_3/2\pi - \omega_2/2\pi$. A plot of the measured electrical reflection as a function of frequency for the interdigital phonon transducers is shown in Supplementary Fig. 5(c). The resonance frequencies are found to be $f_3 = \omega_3/2\pi = 366.8$ MHz, $f_2 = \omega_2/2\pi = 253.5$ MHz, and $f_1 = 113.3$ MHz.

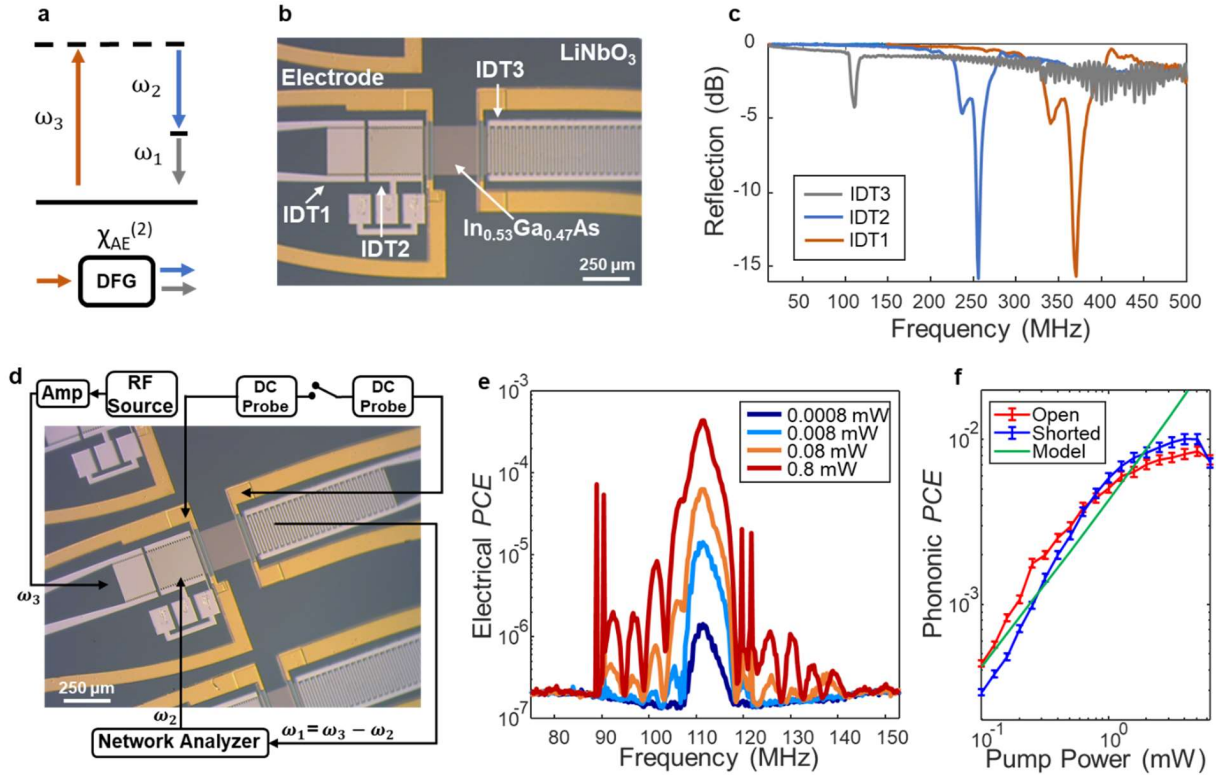

**Supplementary Fig. 5: Difference frequency generation.** (a) Diagram of phonon energy conservation for difference frequency generation. (b) Microscope image of a device to study this parametric three-wave mixing process in the $In_{0.53}Ga_{0.47}As$-$LiNbO_3$ on silicon acoustoelectric heterostructure. The acoustoelectric interaction length, defined by the length of the patterned $In_{0.53}Ga_{0.47}As$, is 250 μm. (c) Measured reflection as a function of frequency for the three interdigital transducers (IDTs) used to generate and detect phonons in the system. (d) A schematic of the experimental setup for power conversion efficiency measurements. (e) Measured electrical power conversion efficiency (*PCE*) as a function of detection frequency for increasing internal pump powers. (f) Phononic power conversion efficiency (*PCE*) as a function of internal pump power for open and shorted electrical contact to the $In_{0.53}Ga_{0.47}As$. The modeled predicted value for the phononic *PCE* is also shown.

A schematic of the experimental setup to measure power conversion efficiency for difference frequency generation is shown in Supplementary Fig. 5(d), and the setup is described in detail in the Methods. As in the sum frequency generation experiments described in the main text, a radio frequency signal generator is used to generate $f_3$ and a network analyzer is used to generate $f_2$ and detect $f_3$. As described in the main text, the electrical conversion efficiency includes the transducer conversion losses and the phononic power conversion efficiency is determined by the electrical power conversion efficiency and measured conversion losses from separate acoustic delay line test structures (Supplementary Note 1). Supplementary Fig. 5(e) shows the measured terminal power conversion efficiency as a function of the detection frequency for increasing internal pump powers. The power conversion has a peak value at the transducer resonance frequency which increases with increasing pump power. The 3 dB fractional bandwidth supported for the frequency conversion measurement is 2.9%, limited by the output transducer bandwidth.

Supplementary Fig. 5(f) shows a plot of the phononic power conversion efficiency as a function of internal pump power for open and shorted electrical contacts to the $In_{0.53}Ga_{0.47}As$ semiconducting layer. A maximum internal power conversion efficiency of $(1.0\pm0.1)\%$ is achieved for the shorted electrode configuration with an internal pump power of 5 mW. The expected theoretical values, as determined by numerical integration of the nonlinear coupled-mode equations (Supplementary Note 3), are also shown in Supplementary Fig. 5(f). Good agreement between the modeled and experimental values occurs up to an internal pump power of approximately 2 mW above which the experimental values are significantly less than the theoretically predicted values. As discussed in the main text, deviation of the experimental values from the theoretical values at large pump power could be due to several effects including pump depletion, self-phase and cross-phase modulation, and thermal detuning.

**Supplementary Note 6**
**Phase matching with acoustic and acoustoelectric dispersion**
For a second-order nonlinear interaction characterized by the relation $\omega_3 = \omega_1 + \omega_2$, the collinear phase-matching condition is $\beta_3 = \beta_1 + \beta_2$, which for the case of phononic nonlinear interactions, is equivalent to $\omega_3/v_{a,3} = \omega_1/v_{a,1} + \omega_2/v_{a,2}$. There are multiple sources of dispersion in our system that lead to the phase mismatch. The $LiNbO_3$ planar waveguide geometry introduces modal dispersion where the speed of sound for a given mode depends on the thickness of the waveguide and the mode's polarization. The acoustic velocity takes a value between the speed of sound in the $LiNbO_3$ and the speed of sound in the silicon for each acoustic wavelength value. The ~50 nm thick $In_{0.53}Ga_{0.47}As$ film modifies this modal dispersion by mass loading effects on top of the $LiNbO_3$. In addition, the same electron-phonon acoustoelectric interaction that leads to linear attenuation of the phonons also modifies the acoustic velocity and is a function of the frequency.[32,47]

From delay line test structures with interdigital transducers designed to launch and detect phonons across a range of frequencies, we can determine the acoustic dispersion for the generated phonons. These test structures do not include the $In_{0.53}Ga_{0.47}As$ film and therefore only provide information on the dispersion due to the $LiNbO_3$ film alone. A plot of the measured acoustic velocity as a function of the frequency is shown in Supplementary Fig. 6(a). Without the $In_{0.53}Ga_{0.47}As$ semiconductor film, our phase mismatch is given by $\Delta\beta = \frac{\omega_3}{v_{a,3}} - \frac{\omega_1}{v_{a,1}} - \frac{\omega_2}{v_{a,2}} = \frac{2\pi \times 626.6 \text{ MHz}}{4043 \frac{m}{s}} - \frac{2\pi \times 370 \text{ MHz}}{4060 \frac{m}{s}} - \frac{2\pi \times 256.6 \text{ MHz}}{4110 \frac{m}{s}} = 89 \text{ cm}^{-1}$ for sum frequency generation and $\Delta\beta = \frac{\omega_3}{v_{a,3}} - \frac{\omega_1}{v_{a,1}} - \frac{\omega_2}{v_{a,2}} = \frac{2\pi \times 367.2 \text{ MHz}}{4060 \frac{m}{s}} - \frac{2\pi \times 113.7 \text{ MHz}}{4415 \frac{m}{s}} - \frac{2\pi \times 253.5 \text{ MHz}}{4110 \frac{m}{s}} = 189 \text{ cm}^{-1}$ for difference frequency generation. These values are taken directly from experimental measurements of acoustic delay line test structures.

A finite element method model was utilized to predict the impact of the mass loading of the $In_{0.53}Ga_{0.47}As$ film on the dispersion. A 50 nm $In_{0.53}Ga_{0.47}As$ film was placed on the top surface of the 5 μm thick $LiNbO_3$, which is placed on top of a silicon substrate. The acoustic frequency of the quasi shear horizontal ($SH_0$) mode was monitored as a function of changing the supported acoustic wavelength for the structure and from this information we determined the acoustic velocity as a function of the acoustic wavelength (Supplementary Fig. 6(b)).

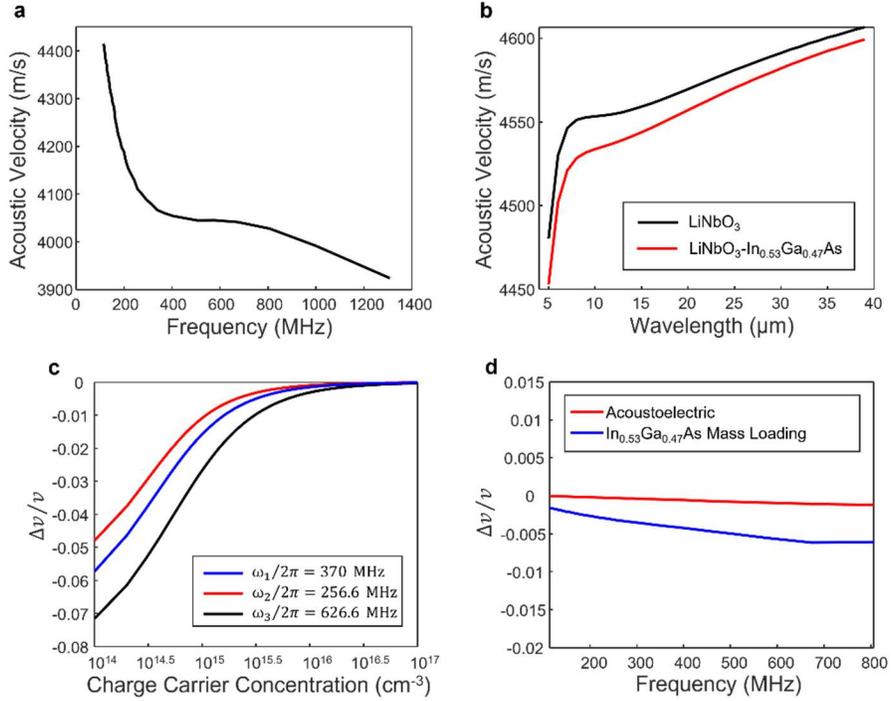

**Supplementary Fig. 6: Dispersion and phase mismatch.** (a) Measured acoustic velocity as a function of frequency for acoustic delay line test structures on the LiNbO$_3$-silicon substrate. (b) Modeled acoustic velocity as a function of wavelength with and without the mass loading effect of the 50 nm thick In$_{0.53}$Ga$_{0.47}$As film. (c) Theoretical calculations of the velocity change ($\Delta v/v$) as a function of semiconductor charge carrier concentration for the pump ($\omega_1/2\pi = 370$ MHz), signal ($\omega_2/2\pi = 256.6$ MHz), and sum frequency ($\omega_3/2\pi = 626.6$ MHz) as a function of charge carrier concentration due to dispersion from the acoustoelectric effect. (d) Expected velocity change $\Delta v/v$ as a function of frequency due to acoustoelectric dispersion and dispersion from the 50 nm thick In$_{0.53}$Ga$_{0.47}$As film.

It is well-known that the acoustoelectric effect itself is dispersive and a model of the acoustoelectric dispersion in a heterostructure like ours has been previously developed[32] which we can apply here to estimate the velocity change. A plot of the theoretical $\Delta v/v$ as a function of semiconductor charge carrier density is shown in Supplementary Fig. 6(d) for the pump frequency ($\omega_1/2\pi = 370$ MHz), the signal frequency ($\omega_2/2\pi = 256.6$ MHz), and the sum frequency ($\omega_3/2\pi = 626.6$ MHz). As can be seen, the dispersion due to the acoustoelectric effect increases with decreasing carrier concentration and increasing frequency for the range of carrier concentration and frequencies shown in Supplementary Fig. 6(c).

A plot of the modeled and theoretical $\Delta v/v$ as a function of frequency is shown in Supplementary Fig. 6(d) for the cases of acoustoelectric dispersion and In$_{0.53}$Ga$_{0.47}$As mass loading. This plot is used to approximate appropriate modifications to the measured acoustic velocities shown in Supplementary Fig. 5(a). Adding these two effects together gives us the total change in acoustic velocity as a function of frequency, which is utilized to adjust the measured values. Including the impact of mass loading and the acoustoelectric effect, the phase mismatch for sum frequency generation is $\Delta\beta = \frac{2\pi \times 626.6 \text{ MHz}}{4016\frac{m}{s}} - \frac{2\pi \times 370 \text{ MHz}}{4041\frac{m}{s}} - \frac{2\pi \times 256.6 \text{ MHz}}{4095\frac{m}{s}} = 113 \text{ cm}^{-1}$.

The phase mismatch for difference frequency generation is approximately $\Delta\beta = \frac{2\pi \times 367.2 \text{ MHz}}{4029\frac{m}{s}} - \frac{2\pi \times 113.7 \text{ MHz}}{4408\frac{m}{s}} - \frac{2\pi \times 253.5 \text{ MHz}}{4092\frac{m}{s}} = 213 \text{ cm}^{-1}$. While the phase mismatch has increased compared to the LiNbO$_3$ film alone, we do see that the majority of the phase mismatch, for the material parameters and frequencies utilized here, can be attributed to the dispersion of the LiNbO$_3$ planar waveguide.

**Supplementary Note 7**
**Linear and nonlinear acoustoelectric coefficients with an applied drift field**
The variation in the linear and nonlinear acoustoelectric coefficients ($\alpha_3$ and $|\eta_3 u_1|$), respectively) with the ratio between the charge carrier drift velocity, $v_d$, and the acoustic velocity, $v_{a,3}$, is shown in Supplementary Fig. 7. A positive velocity ratio corresponds to when the charge carriers drift in the same direction as the propagating piezoelectric phonons while a negative velocity ratio corresponds to when the direction of charge carrier drift is opposite to the phonon propagation. At a velocity ratio equal to 1, the linear acoustoelectric coefficient is zero and the nonlinear acoustoelectric coefficient reaches a minimum value. As the velocity ratio increases from 1, the magnitude of the nonlinear acoustoelectric coefficient increases while negative attenuation (amplification) is achieved. For a negative velocity ratio, attenuation increases, but so does the magnitude of the nonlinear acoustoelectric coefficient.

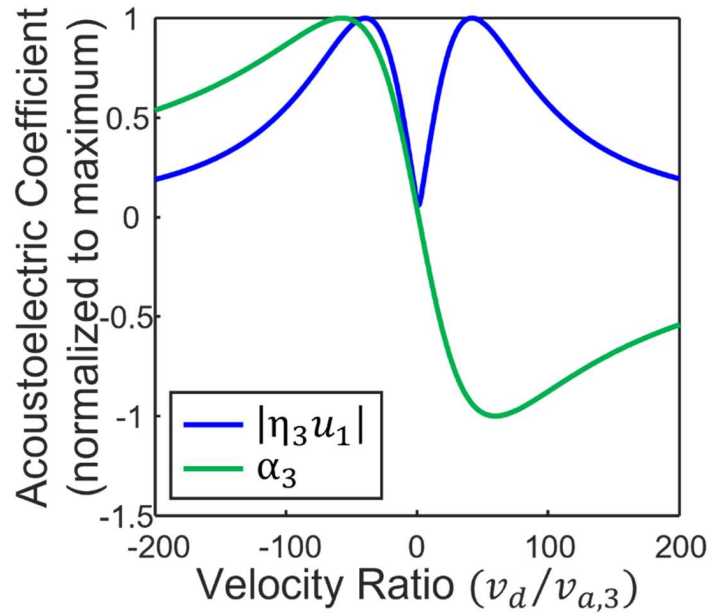

**Supplementary Fig. 7: Dispersion and phase mismatch.** The linear, $\alpha_3$, and nonlinear, $|\eta_3 u_1|$, acoustoelectric coefficients are shown as a function of the ratio between the charge carrier drift veloctiy, $v_d$, and the acoustic velocity, $v_{a,3}$.

**Supplementary Note 8**
**Four-wave mixing coefficient**

The four-wave mixing coefficeint and nonlinear modal coefficients were determined using the previously reported methods.[27] The mode amplitude for the phonon displacement for the $n^{\text{th}}$ waveguide mode generated by the cascaded nonlinearity, $A_n$, is given by

$$\frac{dA_n}{dx} = -\frac{\alpha}{2} A_n + i\gamma_m \sum_{p+q-m=n} A_p A_q A_m^* e^{i\Delta\beta x} \tag{S12}$$

where $\alpha$ is the phononic propagation loss, $\gamma_m$ is the modal nonlinear coefficient, $\Delta\beta = \beta_p + \beta_p - \beta_m - \beta_n$ is the phase mismatch, and $A_n$ is normalized such that the acoustic power in the $n^{\text{th}}$ mode, $P_n$, is $P_n = |A_n|^2$. For the small frequency spacing and interaction lengths, $L$, we study in this work, we can assume that $\Delta\beta_{ij}L$ is approximately equal to 0 for all modes $i, j$ studied. In the low power limit and under the assumption that the phonons in the first sidebands are created by degenerate four-wave mixing, the efficiencies with which these phonons are created is given by

$$PCE_{112} = \left|\frac{A_{f112}}{A_{f2}^*}\right|^2 = \Gamma_{112} P_1'^2 \tag{S13a}$$

$$PCE_{221} = \left|\frac{A_{f221}}{A_{f1}^*}\right|^2 = \Gamma_{221} P_2'^2 \tag{S13b}$$

where $PCE_{112}$ is the power conversion efficiency to generate the phonons with amplitude $A_{f112}$ at frequency $f_{112}$, $PCE_{221}$ is the power conversion efficiency to generate the phonons with amplitude $A_{f221}$ at $f_{221}$, $A_{f1}$ and $A_{f2}$ are the amplitudes for the pumps with frequencies $f_1$ and $f_2$, respectively, $P_1'$ and $P_2'$ are the input pump powers with frequencies $f_1$ and $f_2$, respectively, at the beginning of the waveguide directly after the input interdigital transducer, and $\Gamma = (\gamma_m L_{\text{eff}})^2$ is the FWM coefficient where $L_{\text{eff}} = [1 - \exp(-\alpha L)]/\alpha$ is the effective length and $L$ is the interaction length. Two values for $\Gamma$ are then found by determining the slopes of $P_{112}/P_2$ as a function of $P_1'^2$ and $P_{221}/P_1$ as a function of $P_2'^2$, and we use $\Gamma = (\Gamma_{112} + \Gamma_{221})/2$ to assess the net nonlinearity. The modal nonlinear coefficients, $\gamma_m$, are calculated according to the values, for the different device types, for $\Gamma$ and the values for $L_{\text{eff}}$, where the $L_{\text{eff}}$ values are calculated from the experimentally extracted acoustoelectric attenuation and the acoustic propagation losses.

**References**
1  Hendrickson, S. M., Foster, A. C., Camacho, R. M. & Clader, B. D. Integrated nonlinear photonics: emerging applications and ongoing challenges. *JOSA B* **31**, 3193-3203 (2014).
2  Leuthold, J., Koos, C. & Freude, W. Nonlinear silicon photonics. *Nature photonics* **4**, 535-544 (2010).
3  Moody, G., Chang, L., Steiner, T. J. & Bowers, J. E. Chip-scale nonlinear photonics for quantum light generation. *AVS Quantum Science* **2**, 041702 (2020).
4  Bers, A. & Cafarella, J. Surface state memory in surface acoustoelectric correlator. *Applied Physics Letters* **25**, 133-135 (1974).
5  Cafarella, J. H., Brown, W., Stern, E. & Alusow, J. Acoustoelectric convolvers for programmable matched filtering in spread-spectrum systems. *Proceedings of the IEEE* **64**, 756-759 (1976).

6       Kino, G. S. Acoustoelectric interactions in acoustic-surface-wave devices. *Proceedings of the IEEE* **64**, 724-748 (1976).
7       Reible, S. A. Acoustoelectric convolver technology for spread-spectrum communications. *IEEE Transactions on Microwave Theory and Techniques* **29**, 463-474 (1981).
8       Hackett, L. *et al.* Towards single-chip radiofrequency signal processing via acoustoelectric electron–phonon interactions. *Nature communications* **12**, 2769 (2021).
9       Bogdanov, S., Shalaginov, M., Boltasseva, A. & Shalaev, V. M. Material platforms for integrated quantum photonics. *Optical Materials Express* **7**, 111-132 (2017).
10      Elshaari, A. W., Pernice, W., Srinivasan, K., Benson, O. & Zwiller, V. Hybrid integrated quantum photonic circuits. *Nature Photonics* **14**, 285-298 (2020).
11      Politi, A., Matthews, J. C., Thompson, M. G. & O'Brien, J. L. Integrated quantum photonics. *IEEE Journal of Selected Topics in Quantum Electronics* **15**, 1673-1684 (2009).
12      MacCabe, G. S. *et al.* Nano-acoustic resonator with ultralong phonon lifetime. *Science* **370**, 840-843 (2020).
13      Wollack, E. A. *et al.* Loss channels affecting lithium niobate phononic crystal resonators at cryogenic temperature. *Applied Physics Letters* **118**, 123501 (2021).
14      Guo, Y. & Wang, M. Phonon hydrodynamics and its applications in nanoscale heat transport. *Physics Reports* **595**, 1-44 (2015).
15      Joshi, A. & Majumdar, A. Transient ballistic and diffusive phonon heat transport in thin films. *Journal of Applied Physics* **74**, 31-39 (1993).
16      Arora, V. K. & Naeem, A. Phonon-scattering-limited mobility in a quantum-well heterostructure. *Physical Review B* **31**, 3887 (1985).
17      Hwang, E. & Sarma, S. D. Acoustic phonon scattering limited carrier mobility in two-dimensional extrinsic graphene. *Physical Review B* **77**, 115449 (2008).
18      Zhou, J.-J. & Bernardi, M. Ab initio electron mobility and polar phonon scattering in GaAs. *Physical Review B* **94**, 201201 (2016).
19      Shao, L. *et al.* Electrical control of surface acoustic waves. *Nature Electronics* **5**, 348-355 (2022).
20      Abdelkefi, A., Nayfeh, A. H. & Hajj, M. R. Effects of nonlinear piezoelectric coupling on energy harvesters under direct excitation. *Nonlinear Dynamics* **67**, 1221-1232 (2012).
21      Mahboob, I., Wilmart, Q., Nishiguchi, K., Fujiwara, A. & Yamaguchi, H. Wide-band idler generation in a GaAs electromechanical resonator. *Physical Review B* **84**, 113411 (2011).
22      Luukkala, M. v. & Kino, G. Convolution and time inversion using parametric interactions of acoustic surface waves. *Applied Physics Letters* **18**, 393-394 (1971).
23      Kurosu, M., Hatanaka, D., Onomitsu, K. & Yamaguchi, H. On-chip temporal focusing of elastic waves in a phononic crystal waveguide. *Nature communications* **9**, 1331 (2018).
24      Kurosu, M., Hatanaka, D. & Yamaguchi, H. Mechanical Kerr nonlinearity of wave propagation in an on-chip nanoelectromechanical waveguide. *Physical Review Applied* **13**, 014056 (2020).
25      Maksymov, I. S., Huy Nguyen, B. Q., Pototsky, A. & Suslov, S. Acoustic, phononic, Brillouin light scattering and Faraday wave-based frequency combs: Physical foundations and applications. *Sensors* **22**, 3921 (2022).
26      Mansoorzare, H. & Abdolvand, R. in *2023 IEEE 36th International Conference on Micro Electro Mechanical Systems (MEMS).* 1183-1185 (IEEE).

27  Mayor, F. M. *et al.* Gigahertz phononic integrated circuits on thin-film lithium niobate on sapphire. *Physical Review Applied* **15**, 014039 (2021).
28  Conwell, E. & Ganguly, A. Mixing of acoustic waves in piezoelectric semiconductors. *Physical Review B* **4**, 2535 (1971).
29  Johri, G. & Spector, H. N. Boltzmann-equation approach to nonlinear acoustoelectric interactions in piezoelectric semiconductors. *Physical Review B* **12**, 3215 (1975).
30  Wu, C. C. & Spector, H. N. Ultrasonic harmonic generation in piezoelectric semiconductors. *Journal of Applied Physics* **43**, 2937-2944 (1972).
31  Adler, R. Simple theory of acoustic amplification. *IEEE Transactions on Sonics and Ultrasonics* **18**, 115-118 (1971).
32  Kino, G. & Reeder, T. A normal mode theory for the Rayleigh wave amplifier. *IEEE Transactions on Electron Devices* **18**, 909-920 (1971).
33  Coldren, L. A. & Kino, G. Monolithic acoustic surface‐wave amplifier. *Applied Physics Letters* **18**, 317-319 (1971).
34  Hackett, L. *et al.* Non-reciprocal acoustoelectric microwave amplifiers with net gain and low noise in continuous operation. *Nature Electronics*, 1-10 (2023).
35  Mansoorzare, H. & Abdolvand, R. Micromachined Heterostructured Lamb Mode Waveguides for Acoustoelectric Signal Processing. *IEEE Transactions on Microwave Theory and Techniques* **70**, 5195-5204 (2022).
36  Boyd, R. W. *Nonlinear optics*. (Academic press, 2020).
37  Rodriguez, A., Soljačić, M., Joannopoulos, J. D. & Johnson, S. G. $\chi^{(2)}$ and $\chi^{(3)}$ harmonic generation at a critical power in inhomogeneous doubly resonant cavities. *Optics express* **15**, 7303-7318 (2007).
38  Eichenfield, M. & Olsson, R. in *2013 IEEE International Ultrasonics Symposium (IUS)*. 753-756 (IEEE).
39  Kumar, P. Quantum frequency conversion. *Optics letters* **15**, 1476-1478 (1990).
40  Banszerus, L. *et al.* Ultrahigh-mobility graphene devices from chemical vapor deposition on reusable copper. *Science advances* **1**, e1500222 (2015).
41  Kim, M.-S. *et al.* Sheet resistance analysis of interface-engineered multilayer graphene: mobility versus sheet carrier concentration. *ACS applied materials & interfaces* **12**, 30932-30940 (2020).
42  Laroche, D., Das Sarma, S., Gervais, G., Lilly, M. & Reno, J. Scattering mechanism in modulation-doped shallow two-dimensional electron gases. *Applied Physics Letters* **96**, 162112 (2010).
43  Otterstrom, N. T. *et al.* Nonreciprocal frequency domain beam splitter. *Physical review letters* **127**, 253603 (2021).
44  Mirhosseini, M., Sipahigil, A., Kalaee, M. & Painter, O. Superconducting qubit to optical photon transduction. *Nature* **588**, 599-603 (2020).
45  Jiang, W. *et al.* Efficient bidirectional piezo-optomechanical transduction between microwave and optical frequency. *Nature communications* **11**, 1166 (2020).
46  Auld, B. A. *Acoustic fields and waves in solids*. (Рипол Классик, 1973).
47  White, D. L. Amplification of ultrasonic waves in piezoelectric semiconductors. *Journal of Applied Physics* **33**, 2547-2554 (1962).